\documentclass[superscriptaddress, twocolumn]{revtex4}

\usepackage{graphicx}

\begin{document}

\title{Urban green space and happiness in developed countries}

\author{Oh-Hyun Kwon}
\thanks{These two authors contributed equally.}
\affiliation{Department of Physics, Pohang University of Science and Technology, Pohang 37673, Republic of Korea.}

\author{Inho Hong}
\thanks{These two authors contributed equally.}
\affiliation{Center for Humans and Machines, Max Planck Institute for Human Development, Berlin 14195, Germany.}

\author{Jeasurk Yang}
\affiliation{Department of Geography, National University of Singapore, Singapore 119260, Singapore.}

\author{Donghee Yvette Wohn}
\affiliation{Department of Informatics, New Jersey Institute of Technology, Newark, NJ 07103, USA.}

\author{Woo-Sung Jung}
\email{wsjung@postech.ac.kr}
\affiliation{Department of Physics, Pohang University of Science and Technology, Pohang 37673, Republic of Korea.}
\affiliation{Department of Industrial and Management Engineering, Pohang University of Science and Technology, Pohang 37673, Republic of Korea.}
\affiliation{Asia Pacific Center for Theoretical Physics, Pohang 37673, Republic of Korea.}

\author{Meeyoung Cha}
\email{mcha@ibs.re.kr}
\affiliation{Data Science Group, Institute for Basic Science, Daejeon 34126, Republic of Korea.}
\affiliation{School of Computing, Korea Advanced Institute of Science and Technology, Daejeon 34141, Republic of Korea.}


\begin{abstract}
Urban green space has been regarded as contributing to citizen happiness by promoting physical and mental health. However, how urban green space and happiness are related across many countries of different socioeconomic conditions has not been explained well. By measuring urban green space score (UGS) from high-resolution Sentinel-2 satellite imagery of 90 global cities that in total cover 179,168 km$^2$ and include 230 million people in 60 developed countries, we reveal that the amount of urban green space and the GDP can explain the happiness level of the country. More precisely, urban green space and GDP are each individually associated with happiness; happiness in the 30 wealthiest countries is explained only by urban green space, whereas GDP alone explains happiness in the 30 other countries in this study. Lastly, we further show that the relationship between urban green space and happiness is mediated by social support and that GDP moderates the relationship between social support and happiness, which underlines the importance of maintaining urban green space as a place for social cohesion in promoting people's happiness.
\end{abstract}

\maketitle

\section{Introduction}

The advantages of urban green space for public health and urban planning have been of great interest in recent years. Green spaces such as parks, gardens, street trees, riversides, and even private backyards facilitate physical activity, social events, mental relaxation, and relief from stress and heat, thereby leading to direct and indirect benefits for mental and physical health \cite{deVries2003, Dadvand2016}. Thus, worldwide policy changes and efforts have been made to build more urban green space to create sustainable and comfortable living environments \cite{UN2015}.

Urban green space and happiness are known to have an implicit positive correlation. Although this association is still unclear, five pathways through which greenery might have beneficial effects have been reported: relieving stress, stimulating physical activity, facilitating social interactions, generating aesthetic enjoyment, and facilitating a sense of shelter from and adjustment to environmental stressors \cite{deVries2013,Dadvand2016, Liu2019}. Studies have suggested that the same pathways exist in numerous countries \cite{Dzhambov2018}. Among them, social interaction facilitation has been confirmed with strong evidence. Studies \cite{Maas2009,jennings2019relationship} have shown that open green space promotes social cohesion by providing places for social contact; people can naturally encounter neighbors in local green spaces while walking dogs, gardening, and having outdoor parties, which enhances community engagement. Moreover, larger green areas such as parks can hold larger events and activities, enabling social mixing between communities.

The amount of urban green space can be captured mainly by three kinds of measurements: qualitative ratings of observers \cite{Kweon1998, deVries2013}, national land-use and land-cover database \cite{Maas2006, Mackerron2013, Alcock2014}, and geographic information system (GIS) techniques. Among these measurements, GIS techniques are the most recently developed method. One example is utilizing the normalized difference vegetation index (NDVI), a vegetation index computed from Landsat series satellite images (30 m resolution)~\cite{Beyer2014,Dzhambov2018,Liu2019}. Studies such as by Tsai \textit{et al}. \cite{Tsai2018} introduced multiple landscape metrics based on GIS and showed a strong association between green space and mental health in U.S. metropolitan areas. These studies assume the distance from an individual’s residence to the nearest green space has associations with health data \cite{Stigsdotter2010,Dadvand2016}. The green space level was then measured as the fraction of areas with NDVI values above a certain threshold (e.g., 0.2 to 0.4 for sparse vegetation and 0.6 for highly dense vegetation)~\cite{EOS2019NDVI}. However, this method raises the question of how to set an appropriate NDVI threshold for global cities. 

Despite the rich literature on green space's mental benefits, they still have limitations as global-scale comparative research. First, the analytical settings are based on a limited number of Western countries~\cite{Liu2019}; most of these studies have been conducted in the United States \cite{Beyer2014,Tsai2018} and Europe \cite{Dadvand2016, Dzhambov2018}. Moreover, only a few are based on multi-country settings that enable comparative analysis \cite{vandenBerg2016}. As a result, it is unclear whether the association between green space and mental health is robust in developing countries or only in developed countries. The main limitation arises because there is no global medical dataset providing reliable and standardized mental health surveys from different countries. Moreover, no studies have established which green space measurement is appropriate for analysis across countries. Various methods of measuring green space -- questionnaires, qualitative interviews, satellite images, Google Street View images, and even smartphone technology~\cite{Review2017Markevych} -- still rely on individual-level measurements (e.g., calculating the greenery level around residential buildings) and hence are not scalable to the global level.

This paper presents a new way to analyze the effects of green space on happiness at the planetary scale, incorporate the different countries' different contexts, and achieve robust results. First, we measure the amount of urban green space from high-resolution satellite images for different countries by developing a globally comparable green space metric. Our metric based on the total NDVI of built-up areas enables this comparison as it does not require an arbitrary threshold that varies for different regions. It also overcomes the limitations of official statistics based on national land-use land cover data that tend to have different criteria by countries and often include only official parks and open space. Our analysis on high-resolution (10 m) Sentinel-2 satellite images provides more accurate information of urban green space than the previous studies on the Landsat series images (i.e., the resolution of 30 m)~\cite{Beyer2014, Dadvand2016,Dzhambov2018,Liu2019}.

Next, this study uses selected happiness scores from the World Happiness Report \cite{happiness-report2018}, which provides reliable and standardized data on multiple countries' mental health and allows comparisons among nations. As happiness is a criterion of emotional well-being, it is interconnected with mental health.  From the perspective that economic studies distinguish between emotional well-being (happiness) and life satisfaction (life evaluation)\cite{Kahneman2010income}, we focus on the impact of green space on emotional happiness. Specifically, we study this relationship in the developed countries of the highest Human Development Index (HDI), where green environments in cities are considered more important for well-being. 

Using these datasets from satellite imagery, we explore the relationship between urban green space and happiness globally. Additionally, we identify conditional indirect effects by national wealth and social support by employing a moderated mediation regression model on socioeconomic indicators. 

\section{Urban green space and happiness in countries}

\begin{figure*}[ht]
\includegraphics[width=0.9\linewidth]{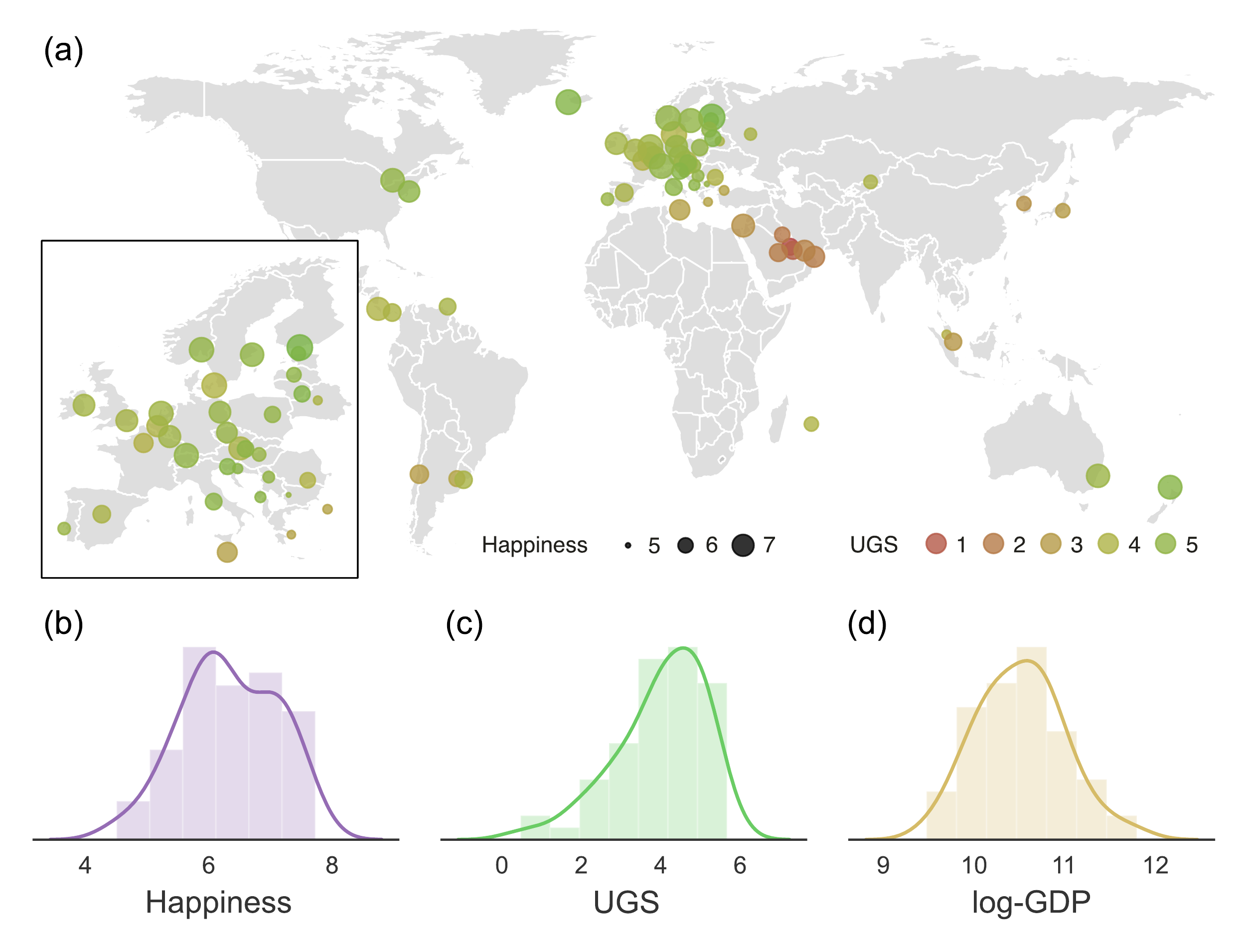}
    \caption{\textbf{The distributions of urban green space and happiness over the world.}~\textbf{(a)} The map of urban green space and happiness in 60 developed countries. The size and color of circles represent the level of happiness and urban green space in a country, respectively. The markers are placed on the most populated cities of each country. \textbf{(b-d)} The histograms of (b) happiness, (c) urban green space (UGS) and (d) logarithmic GDP per capita (log-GDP). We use the logarithm of the total NDVI per capita as an indicator of urban green space, and the logarithm of GDP per capita as a measure of wealth.}
    \label{fig:map}
\end{figure*}

We examine the global relationship between urban green space and happiness in 60 developed countries ranked by the Human Development Index. Using the Sentinel-2 satellite imagery dataset, we define each country's urban green space score (UGS) as a logarithmic total vegetation index per capita in the most populated cities (i.e., those that include at least 10\% of the national population). Among the various vegetation indices available, NDVI \cite{Miura2019NDVI} is used based on the robustness of the results for different tested indices. The happiness score and the gross domestic product based on purchasing power parity (GDP (PPP)) per capita of each country are from the World Happiness Report \cite{happiness-report2018} and the International Monetary Fund (IMF) estimation \cite{IMF2018}, respectively (see the Methods section and the Supplementary Information for details).

Figure~\ref{fig:map}(a) shows an overall view of urban green space and the happiness of countries around the world. This map highlights regional differences in the green space distribution due to climate; countries near the equator in tropical climates have relatively high UGS values, while countries located in the 20-30$^{\circ}$ latitude range have exceptionally low UGS values due to the dry climate. UGS increases with latitude in higher-latitude regions. On the other hand, Northern and Western European and North American countries display relatively large happiness. Western Asian countries also show relatively high happiness with low UGS value, indicating that the relationship between happiness and green space is not trivial.

Figure~\ref{fig:map}(b-d) shows the distribution of happiness, UGS, and log-GDP, and they all show unimodal distributions with low skewness, which is appropriate for linear regression analyses. Note that the probability distributions of NDVI per capita and GDP per capita converge to a normal distribution after logarithmic scaling. Our comparison of several green space measures shows that the logarithmic NDVI per capita is most suitable for the following analysis in terms of its distribution and explanatory power. We hence choose the logarithmic NDVI per capita as the primary green space indicator in this research. (see Supplementary Information). We also use the logarithmic GDP per capita (PPP) (hereinafter referred to as the log-GDP) as a measure of the wealth of the country, as noted in the Happiness Report \cite{happiness-report2018}. 

\begin{table*}[ht]
\begin{tabular}{c|c c c|c c c|c c c}
\hline
Countries & & All & & & Lower 30 & & & Top 30 & \\
\hline
Model & (1) & (2) & (3) & (4) & (5) & (6) & (7) & (8) & (9)\\
\hline
log-GDP & 1.0120\textsuperscript{***} & - & 1.1319\textsuperscript{***} & 0.9034\textsuperscript{**} & - & 0.8517\textsuperscript{*} & -0.0809 & - & 0.2581\\
 & (0.6603) & & (0.6234) & (1.6305) & & (1.7493) & (1.3559) & & (1.0314)\\
UGS & - & 0.1165 & 0.2249\textsuperscript{***} & - & 0.1497 & 0.0567 & - & 0.2785\textsuperscript{***} & 0.2946\textsuperscript{***}\\
 & & (0.3545) & (0.2643) & & (0.6042) & (0.6051) & & (0.2313) & (0.2403) \\
Const & -4.2945\textsuperscript{**} & 5.9007\textsuperscript{***} & -6.4709\textsuperscript{***} & -3.3428 & 5.1767\textsuperscript{***} & -3.0629 & 7.7712\textsuperscript{**} & 5.8110\textsuperscript{***} & 2.9312\\
 & (6.9672) & (1.4910) & (6.8998) & (16.5490) & (2.6490) & (17.1094) & (14.8065) & (0.9455) & (11.5463) \\
\hline
Adjusted $R^2$ & 0.3832 & 0.00123 & 0.4786 & 0.1296 & 0.0012 & 0.1013 & -0.0335 & 0.4457 & 0.4468 \\
Observations & 60 & 60 & 60 & 30 & 30 & 30 & 30 & 30 & 30\\
\hline
\end{tabular}
\caption{\label{tab:reg_gdp}\textbf{The regression analysis for happiness, UGS, and log-GDP.} 
The values denote the regression coefficients and the confidence intervals of each independent variable with its significance (i.e., \textsuperscript{***}p<0.01; \textsuperscript{**}p<0.05; \textsuperscript{*}p<0.1). The regression model (1-3), model (4-6) and model (7-9) are examined for the data of all countries, the lower 30 countries and the top 30 countries ranked by GDP, respectively.}
\end{table*}

As per-country wealth is an important indicator of its citizens' quality of life, wealth (i.e., log-GDP) should be considered in analyzing urban green space and happiness. Our regression analysis finds that UGS, together with log-GDP, explains happiness. We make new observations from Table~\ref{tab:reg_gdp}. Although UGS is not substantially correlated with happiness in the simple linear regression (i.e., model (2)), the multilinear model with log-GDP (i.e., model (3)) has a substantial increase in prediction ability compared to the simple regression analysis on log-GDP (i.e., model (1)). Therefore, urban green space adds explanatory power to the correlation between wealth and happiness across countries. The regression analyses with other green space-variant measures further confirm this result's robustness, confirming a substantial increase in the adjusted R-squared value when including UGS in the regression. Specifically, UGS based on the logarithmic NDVI per capita shows the best regression performance (see the Supplementary Information for the results for the different measures).

\section{Urban green space is effective in rich countries}

Our results show that happiness is correlated with urban green space and the GDP of a country. But, is this green space-happiness effect uniform across all countries? Previous studies on the marginal effect of income on happiness suggest that happiness may have a nonlinear relationship with GDP, presumably showing saturation after a specific GDP --- a concept known as the Easterlin paradox \cite{Easterlin2010}. This paradox tells us that increases in happiness through GDP reach a saturation point, yet what factors promote happiness beyond the saturation point is unknown.

\begin{figure*}[ht]
\includegraphics[width=0.9\linewidth]{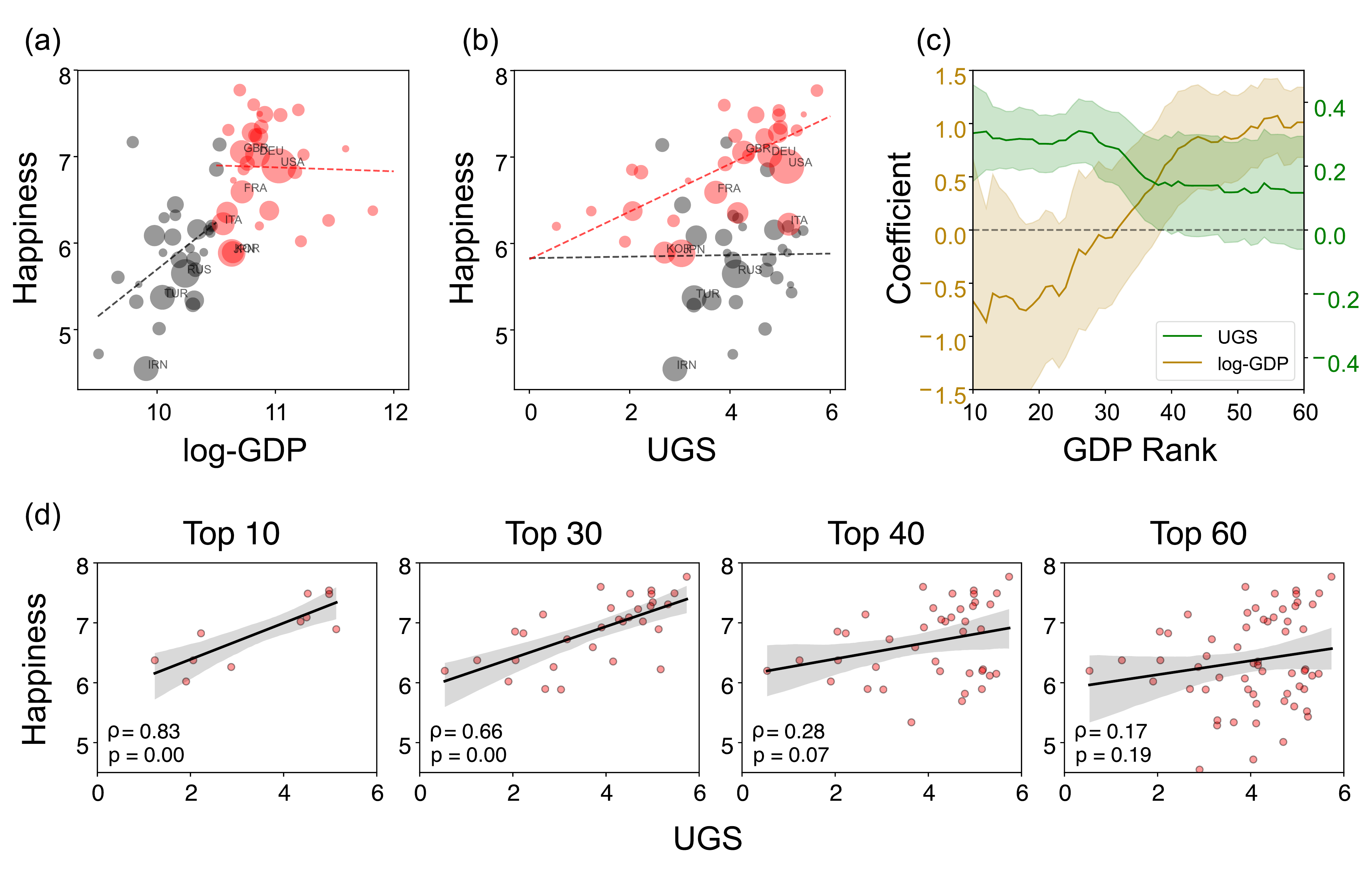}
    \caption{\textbf{The effect of GDP on the green-happiness relation.}~\textbf{(a, b)} The relations of (a) log-GDP and happiness, and (b) urban green space (i.e., UGS) and happiness across 60 developed countries. The top 30 and the lower 30 countries ranked by GDP are sized by the population size and colored by red and black. The dotted lines are the linear fit for each GDP group. \textbf{(c)} Changes of coefficients between urban green space and happiness for different sets of GDP rank with increasing window size from top 10 to 60. \textbf{(d)} The rank correlations between UGS and happiness for the groups of increasing countries in the GDP rank order.}
    \label{fig:gdp}
\end{figure*}

To test the Easterlin paradox, we repeated the analysis over clusters of countries grouped by GDP. Figure~\ref{fig:gdp}(a) shows a high correlation between GDP and happiness in the 30 lower-GDP countries (i.e., $\rho = 0.40$), whereas the correlation is no longer evident in the 30 higher-GDP countries (i.e., $\rho = - 0.04$). These results suggest that economic prosperity (as measured by GDP) is crucial for people's happiness but fails to further promote happiness in rich countries. The GDP appears to reach a happiness-correlation threshold around the 30th wealthiest country, which corresponds to a GDP of 38,518 dollars. Previous research on the Easterlin paradox has stated that the GDP per capita can increase happiness until it reaches a certain threshold but cannot further increase happiness above that threshold. We observe a similar pattern for wealth and happiness across countries. On the other hand, happiness in the 30 wealthiest countries is well explained by urban green space. As shown in Figure~\ref{fig:gdp}(b), urban green space is positively correlated with happiness in the richest countries (i.e., $\rho = 0.66, p < 0.01$), but this correlation is not significant in the 30 lower-GDP countries (i.e., $\rho = 0.19, p = 0.32$). Thus, urban green space is a factor that further increases the happiness of a country after its GDP reaches a certain level.

The regressions for each of the 60 countries ranked by GDP in Table~\ref{tab:reg_gdp} confirm the individual effects of urban green space and GDP on happiness. GDP is the only substantial factor explaining happiness in the 30 lower-GDP countries (models 4-6). In contrast, for the 30 higher-GDP countries, happiness is explained only by the UGS (7-9). These findings suggest that GDP is critical for happiness until it reaches a certain GDP threshold (i.e., the Easterlin paradox), after which urban green space explains happiness better.

The correlation between UGS and happiness also corroborates the effect of UGS in rich countries. The correlation in Figure~\ref{fig:gdp}(d) decreases as more countries in the decreasing order of GDP are added. The correlation is substantial (i.e., $\rho$ is approximately 0.8) among the countries excluding the top 30. Figure~\ref{fig:gdp}(c) summarizes the effects of urban green space and GDP that cross over each other around the 30th wealthiest country. For the top 30 countries, urban green space has positive coefficients, but the GDP effect is not significant. These relationships are reversed for less affluent countries.

In summary, economic support seems to promote happiness until the essential requirements and living standards are met. However, economic success alone fails to add persistent promotion of happiness. After some level, urban green space appears to be related to other social factors that can further promote happiness.

\section{Urban green space for social cohesion}

\begin{figure*}[ht]
\includegraphics[width=0.9\linewidth]{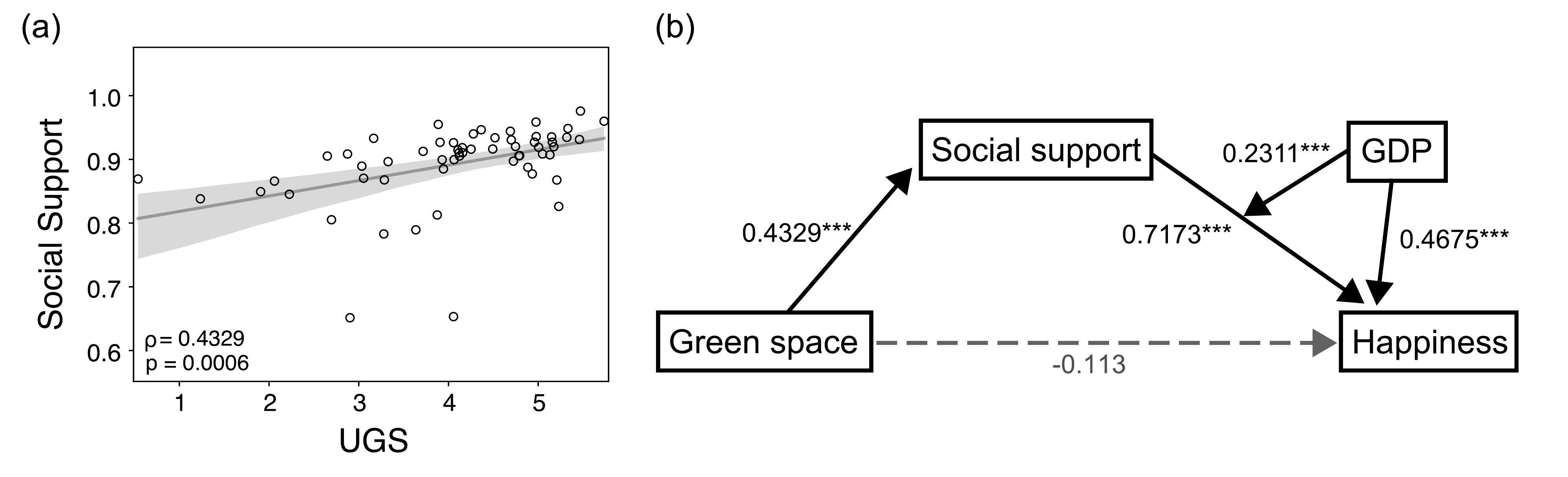}
    \caption{\textbf{The moderated mediation model for UGS, happiness and socioeconomic indicators.}~\textbf{(a)} Scatter plot of social support and UGS across countries. \textbf{(b)} Diagram for the moderated mediation model. The boxes denote the model variables. Solid black arrows denote a statistically significant relationship between a pair of variables with the regression coefficient and the p-value (i.e., ***p<0.01). The gray dashed arrow represents a non-significant relationship. Note that the coefficients are calculated with z-scores of the variables to compare the effect size directly.}
    \label{fig:mod_med}
\end{figure*}

Our findings highlight urban green space as an indicator that might be correlated with social factors promoting happiness beyond the achievement of economic success. The question then arises, which social factors connect urban green space with happiness? To identify this connection, we first examine the correlation between UGS and socioeconomic variables reported in the World Happiness Report: GDP per capita, social support, life expectancy, freedom, generosity, and perceptions of corruption. Of these six variables, only ``social support'' has a significantly positive correlation ($\rho = 0.43, p < 0.01$) with UGS as we can see from Fig. \ref{fig:mod_med}(a), implying that social support could mediate between urban green space and happiness. This relationship is consistent with several existing studies that suggested urban green space as a place of social cohesion \cite{Maas2009,jennings2019relationship}. On the other hand, as indicated by life expectancy, physical health does not display a significant relationship with green space ( $\rho = 0.19, p = 0.15$), contradicting common sense. The regression analysis on happiness with UGS and six socioeconomic variables also captures the interchangeability of urban green space and social support (see the Supplementary Information for details).

Here, we employ a moderated mediation model~\cite{Preacher2007} to characterize the complicated relationships among urban green space, social support, GDP, and happiness. In moderated mediation models, the moderator describes a variable's conditional effect through the interaction term, and the mediator describes a variable's indirect effect connecting the other two variables. Accordingly, moderated mediation models determine the pathway of directed interactions between multiple variables.

First, we examine the mediation effect of urban green space and social support as independent variables. The mediation regression model shows that social support mediates the relationship between urban green space and happiness such that (1) urban green space improves social support and (2) social support promotes happiness. The mediation effect is significant (3) only when GDP is considered in the model. Consequently, our moderated mediation model combining these three effects, shown in Fig. \ref{fig:mod_med}(b), presents the pathway by which green space affects happiness through social support, given that GDP moderates the effect of social support. If we describe this relationship in equations,
\begin{equation}
    H = \beta_0 + \beta_1 M + \beta_2 S + \beta_3 SM,
    \label{eq:moderatedMediation1}
\end{equation}
\begin{equation}
    S = \beta_4 + \beta_5 \ln{G}.
    \label{eq:moderatedMediation2}
\end{equation}
where $H$, $M$, $S$, and $G$ represent happiness, log-GDP, social support, and NDVI per capita, respectively, and the $\beta$ values denote the coefficients of the regression models (see the Supplementary Information for details).

Our moderated mediation model can be used to estimate the amount of urban green space required to increase happiness by a certain amount according to
\begin{equation}
    \Delta H = \left( \beta_1'+\beta_2'M \right)\ln{\frac{G_f}{G_i}}.
    \label{eq:happiness-green}
\end{equation}
In the equation \ref{eq:happiness-green}, the required ratio of urban green spaces in a country decreases as its log-GDP increases. The required increase in urban green space per capita can be estimated for each country based on its current GDP value. For example, the United States needs an additional 36.1908 NDVI of urban green space per capita to increase its happiness score by 0.0546. In contrast, 3,416 USD per capita is required to achieve the same increment in happiness. Here, we used a 0.0546 happiness score as a reference value of $\Delta H$, which is the average value between happiness ranks. Note that the NDVI per capita is interpreted as a weighted area of green space, with a unit of $m^2$. Similarly, Qatar needs 0.4981 NDVI per capita or 7,556 dollars per capita, and South Korea needs 4.1332 NDVI per capita or 2,315 dollars per capita to achieve the reference happiness score increase.

\section{Discussion}

This paper revealed a global relationship between urban green space and happiness in over 60 countries using high-resolution satellite imagery. Urban green space has a higher impact in developed countries (i.e., countries with higher GDPs), which suggests urban green space as a key to promoting happiness beyond economic success. Our moderated mediation model further elucidates this relationship as social support mediates the green-happiness relation, and GDP moderates social support and happiness. This sophisticated model could estimate additional green space needed to promote happiness for each country.

The current study newly defined the concept of UGS (urban green space score), which can be used to calculate the amount of green space at any spatial scale accounting for population density. We compared several green space measures and proposed to use the logarithmic NDVI per capita as a preferred measure of UGS. This index was validated through experiments and it makes it possible to investigate green space at a global level, allowing us to perform cross-sectional research on green space. Furthermore, the method obtaining UGS can be utilized to investigate any spatial areas such as blue space (i.e., aquatic environments such as lake and shore)~\cite{Foley2015, Raymond2016}.  

Our findings have multiple policy-level implications. First, public green space should be made accessible to urban dwellers to enhance social support. In doing so, one critical aspect is public safety. If public safety in urban parks is not guaranteed~\cite{groff2012role,han2018violent}, its positive role in social support and happiness may diminish. The meaning of public safety may change; for example, ensuring biological safety will be a priority in keeping the urban parks accessible during the COVID-19 pandemic \cite{Ugolini2020}.In fact, the high indoor transmission rate of the virus~\cite{Lolli2020} will increase awareness and importance of open spaces like urban parks. While some urban parks may be closed during lockdowns, some reports suggest that viewing them from home could also help relax stress during the pandemic~\cite{Hedblom2019}. Second, urban planning of public green space is needed for both developed and developing countries. While our findings confirmed a strong impact of urban green space on happiness in developed countries, the same positive effect holds for developing countries, albeit to a smaller degree. Furthermore, it is challenging or nearly impossible to secure land for green space after built-up areas are developed in cities. Therefore, urban planning for parks and green recovery (new greening in built-up areas) should be considered in developing economies where new cities and suburban areas rapidly expand \cite{Ewing2008sprawl, Liu2020greenrecovery}.  

In addition to the above, recent climate changes can create substantial volatility in sustaining urban green space. Extreme events such as wildfires, floods, droughts, and cold waves could endanger urban forests around the world \cite{Allen2010climatechange}. On the other hand, global warming could also accelerate tree growth in cities more than in rural areas due to the urban heat island effect \cite{Pretzsch2017climatechange}. In the end, the environmental influence is bidirectional; urban green spaces affect local climates by reducing carbon dioxide levels \cite{Nowak2013carbon} and providing a cooling effect inside the city that indirectly affects people's well-being. Thus, we need more attention to predicting climate changes and discovering their impact on public places since the extreme changes could hamper the benefits of urban green space.

As an exciting future direction, satellite images of higher spatiotemporal resolutions can be used to compute urban green space scores. This paper focused on the correlation across countries fixed in time, given the short span of the Sentinel-2 dataset launched in 2015. A causal analysis could be done with satellite imagery data for a longer span. Also, our dataset does not cover all the countries in the world. Fortunately, our observations from the 30 lower-income countries anticipate the substantial effect of GDP in other developing countries excluded in our analysis. We have analyzed the highest-resolution public dataset of satellite imagery in this study. However, our method still has room for application to higher-resolution non-public datasets such as the household level (less than 10m resolution) available in the national-scale health dataset~\cite{houlden2017cross}. Since satellite imagery cannot account for green space inside buildings (such as green walls), future research could quantify the effect of these mini-scale green spaces using computer vision~\cite{Seiferling2017}.

\section{Methods}

\subsection{Collecting happiness and remote sensing data}

\begin{figure*}[ht]
\includegraphics[width=0.9\linewidth]{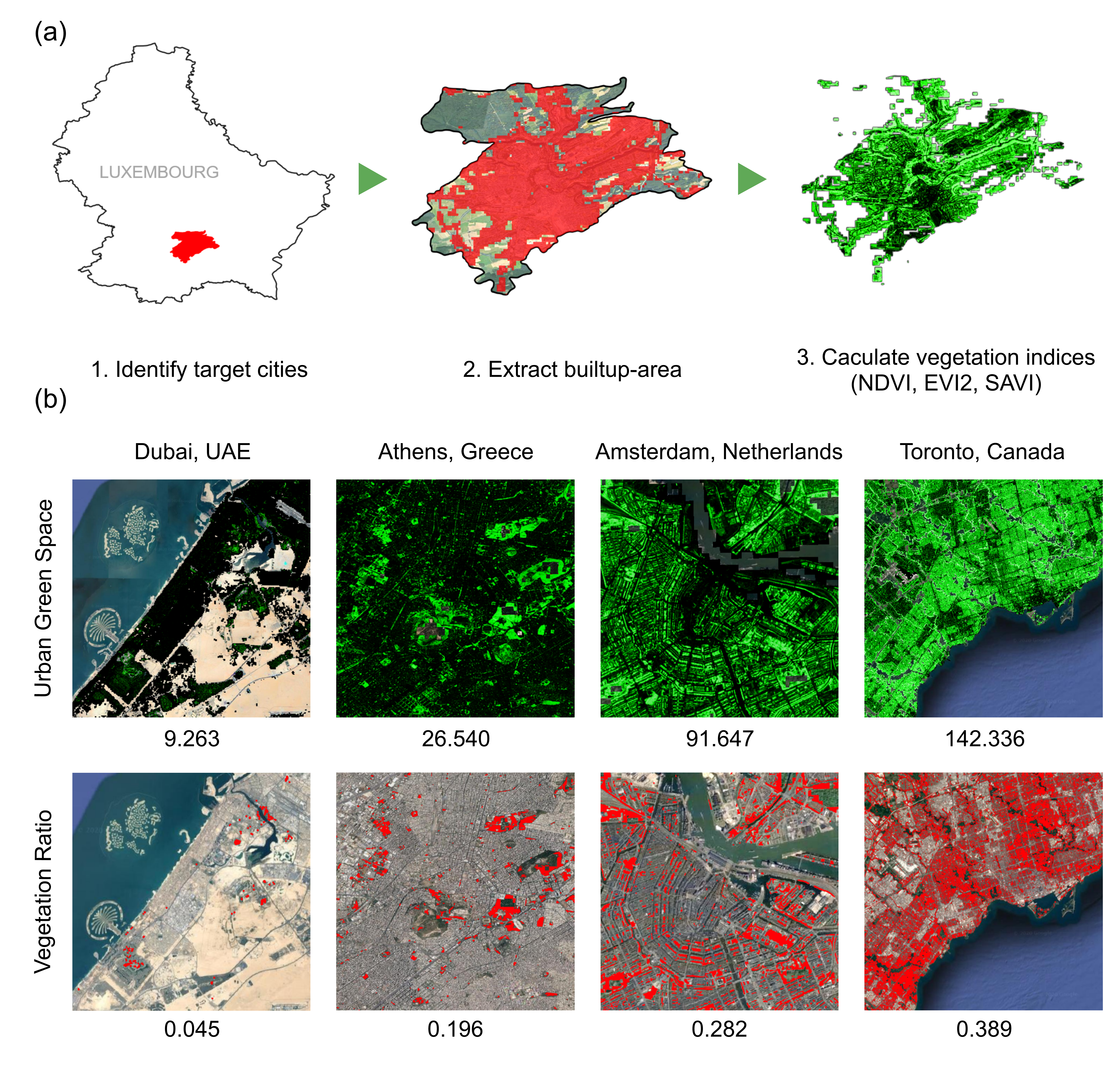}
    \caption{\textbf{Measuring urban green spaces.}~\textbf{(a)} Measurement methods to compute the size of urban green space in each country. First, we find cities occupying more than 10 percent of the total population in a country. Then, we extract the built-up area of the cities with Copernicus global landcover data. Finally, we calculate vegetation indices (e.g., NDVI) within the area using Sentinel-2 satellite images.\textbf{(b)} Urban green space measured by the UGS (upper row) and the vegetation ratio (lower row) in four world cities.The red areas in the upper row indicate vegetation for the NDVI threshold of 0.4. The lower row images show the adjusted NDVI per capita (i.e., UGS) for every 10m by 10m pixel.} 
    \label{fig:Methods}
\end{figure*}

To identify the relationship between happiness and green space, we use happiness scores from the World Happiness Report \cite{happiness-report2018} and the NDVI scores from Sentinel-2 satellite imagery as remote sensing data. The World Happiness Report from 2018 covered 156 countries. The report provides an annual survey of how happy citizens perceive themselves to be and ranks the countries by \textit{happiness score}. The score is the average of the participants' responses asked to rate how happy they are on a scale from 0 and 10. While many socioeconomic indicators (e.g., unemployment and inequality) may affect happiness, not all of these factors are measured annually across 156 countries. The report instead describes happiness with six primary socioeconomic indicators: GDP per capita, social support, life expectancy, freedom to make life choices, generosity, and perceptions of corruption. For example, the social support variable is based on binary responses (yes/no) on a Gallup World Poll question: "If you were in trouble, do you have relatives or friends you can count on to help you whenever you need them, or not?" 

To quantify urban green space in global cities, we use the Sentinel-2 dataset that provides the highest spatial resolution (10 m) among the publicly available satellite imagery datasets (e.g., 30 m resolution in Landsat series) \cite{Beyer2014, Dadvand2016,Dzhambov2018, Liu2019}.
With this high resolution, we can identify granular green space, including street vegetation and home gardens that could not be detected in other public datasets. When using satellite imagery to detect small vegetation, it is critical to consider the season in which the images were obtained \cite{Beyer2014, Dadvand2016, Dzhambov2018, Liu2019}. We use the images from summer: June to September 2018 for the Northern Hemisphere and December 2017 to February 2018 for the Southern Hemisphere. Satellite images with below 10\% cloud cover were used; when such images could not be obtained for the study period, data from 2019 were used instead.

Normalized difference vegetation index (NDVI) is a well-known remote sensing indicator of green vegetation areas in satellite images \cite{Miura2019NDVI}. It detects vegetation as the difference between near-infrared and red light, in the value range from -1 to +1. In general, high NDVI scores include urban green spaces such as official parks, backyards, street trees, mountains, riverbanks, golf courses, and urban farmlands. There are a few well-known variants of NDVI \cite{Review2017Markevych}, such as the soil-adjusted vegetation index (SAVI) \cite{Huete1988SAVI}, which is corrected for soil brightness, and the enhanced vegetation index (EVI) \cite{Jiang2008EVI2}, which is corrected for atmospheric effects. All NDVI, SAVI, and EVI2 scores can be calculated from the two spectral bands of Sentinel-2, red (band 4) and near-infrared (NIR, band 8), as follows:
\begin{equation}
    NDVI = \frac{NIR - RED}{NIR + RED},
\end{equation}
\begin{equation}
    SAVI = \frac{(1 + L)(NIR - RED)}{NIR + RED + L},
\end{equation}
\begin{equation}
    EVI2 = 2.5\frac{NIR - RED}{NIR + 2.4 RED + 1}.
\end{equation}
The robustness of the results for the three green space measures was verified using NDVI as the primary metric.\\

\subsection{Measuring the amount of green space}

The vegetation indices are measured in three steps, as illustrated in Fig. \ref{fig:Methods}(a). The first step is to identify target cities containing at least 10\% of each country's total population and represent the country's overall happiness. The second step is to extract only the built-up areas within the identified cities' administrative boundaries. As cities' boundaries are historically and culturally constructed and often arbitrary, the cities' size needs to be standardized; some cities include vast suburban areas (e.g., Istanbul) or natural areas (e.g., deserts in Dubai). Thus, referring to the global land cover data from the EU's Copernicus Programme~\cite{Copernicus2020}, we focus on urban built-up areas to quantify the urban green space. Finally, the vegetation indices (NDVI, EVI2, and SAVI) are calculated for the extracted urban areas.

The final step is to compute the amount of green space in each country, determined from the measured vegetation indices. Here, we define the amount of green space as the logarithm of the total NDVI of built-up areas in the target cities divided by the cities' total population, called UGS, as a metric for urban green space. UGS is calculated as follows:
\begin{equation}
    UGS = \log{\left(\frac{\sum_c\sum_{b(c)}NDVI(b)}{\sum_{c}N_{c}}\right)},
\end{equation}
where $NDVI(b)$ is the value of NDVI of pixel $b$ within built-up areas $b(c)$ in city $c$ and $N_{c}$ is the population of city $c$. In this calculation, we adjusted negative NDVI values to zero \cite{Review2017Markevych} to prevent errors caused by the accumulation of negative values in areas next to bodies of water (see the Supplementary Information for the entire dataset).

\section*{Acknowledgements}
The authors thank to Farnoosh Hashemi, Ali Behrouz, and Taekho You for useful comments. M.Cha work was supported by the Institute for Basic Science (IBS-R029-C2). 

\section*{Author contributions statement}
M.C. and D.Y.W. conceived the research, I.H., W.-S.J. and M.C. designed the research, O.-H.K. and J.Y. collected the data, O.-H.K. performed the research, O.-H.K., I.H. and M.C. analysed the data, O.-H.K., I.H. and J.Y. wrote the manuscript. All authors reviewed the manuscript. 

\bibliography{ms}

\begin{thebibliography}{41}
\expandafter\ifx\csname natexlab\endcsname\relax\def\natexlab#1{#1}\fi
\expandafter\ifx\csname bibnamefont\endcsname\relax
  \def\bibnamefont#1{#1}\fi
\expandafter\ifx\csname bibfnamefont\endcsname\relax
  \def\bibfnamefont#1{#1}\fi
\expandafter\ifx\csname citenamefont\endcsname\relax
  \def\citenamefont#1{#1}\fi
\expandafter\ifx\csname url\endcsname\relax
  \def\url#1{\texttt{#1}}\fi
\expandafter\ifx\csname urlprefix\endcsname\relax\def\urlprefix{URL }\fi
\providecommand{\bibinfo}[2]{#2}
\providecommand{\eprint}[2][]{\url{#2}}

\bibitem[{\citenamefont{de~Vries et~al.}(2003)\citenamefont{de~Vries, Verheij,
  Groenewegen, and Spreeuwenberg}}]{deVries2003}
\bibinfo{author}{\bibfnamefont{S.}~\bibnamefont{de~Vries}},
  \bibinfo{author}{\bibfnamefont{R.~A.} \bibnamefont{Verheij}},
  \bibinfo{author}{\bibfnamefont{P.~P.} \bibnamefont{Groenewegen}},
  \bibnamefont{and}
  \bibinfo{author}{\bibfnamefont{P.}~\bibnamefont{Spreeuwenberg}},
  \bibinfo{journal}{Environment and Planning A: Economy and Space}
  \textbf{\bibinfo{volume}{35}}, \bibinfo{pages}{1717} (\bibinfo{year}{2003}).

\bibitem[{\citenamefont{Dadvand et~al.}(2016)\citenamefont{Dadvand, Bartoll,
  Basaga\~{n}a, Dalmau-Bueno, Martinez, Ambros, Cirach, Triguero-Mas, Gascon,
  Borrell et~al.}}]{Dadvand2016}
\bibinfo{author}{\bibfnamefont{P.}~\bibnamefont{Dadvand}},
  \bibinfo{author}{\bibfnamefont{X.}~\bibnamefont{Bartoll}},
  \bibinfo{author}{\bibfnamefont{X.}~\bibnamefont{Basaga\~{n}a}},
  \bibinfo{author}{\bibfnamefont{A.}~\bibnamefont{Dalmau-Bueno}},
  \bibinfo{author}{\bibfnamefont{D.}~\bibnamefont{Martinez}},
  \bibinfo{author}{\bibfnamefont{A.}~\bibnamefont{Ambros}},
  \bibinfo{author}{\bibfnamefont{M.}~\bibnamefont{Cirach}},
  \bibinfo{author}{\bibfnamefont{M.}~\bibnamefont{Triguero-Mas}},
  \bibinfo{author}{\bibfnamefont{M.}~\bibnamefont{Gascon}},
  \bibinfo{author}{\bibfnamefont{C.}~\bibnamefont{Borrell}},
  \bibnamefont{et~al.}, \bibinfo{journal}{Environmental International}
  \textbf{\bibinfo{volume}{91}}, \bibinfo{pages}{161} (\bibinfo{year}{2016}).

\bibitem[{\citenamefont{UN}(2015)}]{UN2015}
\bibinfo{author}{\bibnamefont{UN}}, \emph{\bibinfo{title}{Sustainable
  Development Goals}} (\bibinfo{year}{2015}), \bibinfo{note}{available at
  \url{https://sdgs.un.org/goals}. Date accessed 4 November 2020}.

\bibitem[{\citenamefont{de~Vries et~al.}(2013)\citenamefont{de~Vries, van
  Dillen, Groenewegen, and Spreeuwenberg}}]{deVries2013}
\bibinfo{author}{\bibfnamefont{S.}~\bibnamefont{de~Vries}},
  \bibinfo{author}{\bibfnamefont{S.~M.~E.} \bibnamefont{van Dillen}},
  \bibinfo{author}{\bibfnamefont{P.~P.} \bibnamefont{Groenewegen}},
  \bibnamefont{and}
  \bibinfo{author}{\bibfnamefont{P.}~\bibnamefont{Spreeuwenberg}},
  \bibinfo{journal}{Social Science \& Medicine} \textbf{\bibinfo{volume}{94}},
  \bibinfo{pages}{26 } (\bibinfo{year}{2013}).

\bibitem[{\citenamefont{Liu et~al.}(2019)\citenamefont{Liu, Wang, Grekousis,
  Liu, Yuan, and Li}}]{Liu2019}
\bibinfo{author}{\bibfnamefont{Y.}~\bibnamefont{Liu}},
  \bibinfo{author}{\bibfnamefont{R.}~\bibnamefont{Wang}},
  \bibinfo{author}{\bibfnamefont{G.}~\bibnamefont{Grekousis}},
  \bibinfo{author}{\bibfnamefont{Y.}~\bibnamefont{Liu}},
  \bibinfo{author}{\bibfnamefont{Y.}~\bibnamefont{Yuan}}, \bibnamefont{and}
  \bibinfo{author}{\bibfnamefont{Z.}~\bibnamefont{Li}},
  \bibinfo{journal}{Landsacpe and Urban Planning}
  \textbf{\bibinfo{volume}{190}} (\bibinfo{year}{2019}).

\bibitem[{\citenamefont{Dzhambov et~al.}(2018)\citenamefont{Dzhambov, Hartig,
  Markevych, Tilov, and Dimitrova}}]{Dzhambov2018}
\bibinfo{author}{\bibfnamefont{A.}~\bibnamefont{Dzhambov}},
  \bibinfo{author}{\bibfnamefont{T.}~\bibnamefont{Hartig}},
  \bibinfo{author}{\bibfnamefont{I.}~\bibnamefont{Markevych}},
  \bibinfo{author}{\bibfnamefont{B.}~\bibnamefont{Tilov}}, \bibnamefont{and}
  \bibinfo{author}{\bibfnamefont{D.}~\bibnamefont{Dimitrova}},
  \bibinfo{journal}{Environmental Research} \textbf{\bibinfo{volume}{160}},
  \bibinfo{pages}{47 } (\bibinfo{year}{2018}).

\bibitem[{\citenamefont{Maas et~al.}(2009)\citenamefont{Maas, van Dillen,
  Verheij, and Groenewegen}}]{Maas2009}
\bibinfo{author}{\bibfnamefont{J.}~\bibnamefont{Maas}},
  \bibinfo{author}{\bibfnamefont{S.~M.~E.} \bibnamefont{van Dillen}},
  \bibinfo{author}{\bibfnamefont{R.~A.} \bibnamefont{Verheij}},
  \bibnamefont{and} \bibinfo{author}{\bibfnamefont{P.~P.}
  \bibnamefont{Groenewegen}}, \bibinfo{journal}{Health \& Place}
  \textbf{\bibinfo{volume}{15}}, \bibinfo{pages}{586} (\bibinfo{year}{2009}).

\bibitem[{\citenamefont{Jennings and Bamkole}(2019)}]{jennings2019relationship}
\bibinfo{author}{\bibfnamefont{V.}~\bibnamefont{Jennings}} \bibnamefont{and}
  \bibinfo{author}{\bibfnamefont{O.}~\bibnamefont{Bamkole}},
  \bibinfo{journal}{International journal of environmental research and public
  health} \textbf{\bibinfo{volume}{16}}, \bibinfo{pages}{452}
  (\bibinfo{year}{2019}).

\bibitem[{\citenamefont{Kweon et~al.}(1998)\citenamefont{Kweon, Sullivan, and
  Wiley}}]{Kweon1998}
\bibinfo{author}{\bibfnamefont{B.~S.} \bibnamefont{Kweon}},
  \bibinfo{author}{\bibfnamefont{W.~C.} \bibnamefont{Sullivan}},
  \bibnamefont{and} \bibinfo{author}{\bibfnamefont{A.~R.} \bibnamefont{Wiley}},
  \bibinfo{journal}{Environment and Behavior} \textbf{\bibinfo{volume}{30}},
  \bibinfo{pages}{832} (\bibinfo{year}{1998}).

\bibitem[{\citenamefont{Maas et~al.}(2006)\citenamefont{Maas, Verheij,
  Groenewegen, de~Vries, and Spreeuwenberg}}]{Maas2006}
\bibinfo{author}{\bibfnamefont{J.}~\bibnamefont{Maas}},
  \bibinfo{author}{\bibfnamefont{R.~A.} \bibnamefont{Verheij}},
  \bibinfo{author}{\bibfnamefont{P.~P.} \bibnamefont{Groenewegen}},
  \bibinfo{author}{\bibfnamefont{S.}~\bibnamefont{de~Vries}}, \bibnamefont{and}
  \bibinfo{author}{\bibfnamefont{P.}~\bibnamefont{Spreeuwenberg}},
  \bibinfo{journal}{Journal of Epidemiology \& Community Health}
  \textbf{\bibinfo{volume}{60}}, \bibinfo{pages}{587} (\bibinfo{year}{2006}).

\bibitem[{\citenamefont{MacKerron and Mourato}(2013)}]{Mackerron2013}
\bibinfo{author}{\bibfnamefont{G.}~\bibnamefont{MacKerron}} \bibnamefont{and}
  \bibinfo{author}{\bibfnamefont{S.}~\bibnamefont{Mourato}},
  \bibinfo{journal}{Global Environmental change} \textbf{\bibinfo{volume}{23}},
  \bibinfo{pages}{992 } (\bibinfo{year}{2013}).

\bibitem[{\citenamefont{Alcock et~al.}(2014)\citenamefont{Alcock, White,
  Wheeler, Fleming, and Depledge}}]{Alcock2014}
\bibinfo{author}{\bibfnamefont{I.}~\bibnamefont{Alcock}},
  \bibinfo{author}{\bibfnamefont{M.~P.} \bibnamefont{White}},
  \bibinfo{author}{\bibfnamefont{B.~W.} \bibnamefont{Wheeler}},
  \bibinfo{author}{\bibfnamefont{L.~E.} \bibnamefont{Fleming}},
  \bibnamefont{and} \bibinfo{author}{\bibfnamefont{M.~H.}
  \bibnamefont{Depledge}}, \bibinfo{journal}{Environmental Science \&
  Technology} \textbf{\bibinfo{volume}{48}}, \bibinfo{pages}{1247 }
  (\bibinfo{year}{2014}).

\bibitem[{\citenamefont{Beyer et~al.}(2014)\citenamefont{Beyer, Kaltenbach,
  Szabo, Bogar, Nieto, and Malecki}}]{Beyer2014}
\bibinfo{author}{\bibfnamefont{K.~M.~M.} \bibnamefont{Beyer}},
  \bibinfo{author}{\bibfnamefont{A.}~\bibnamefont{Kaltenbach}},
  \bibinfo{author}{\bibfnamefont{A.}~\bibnamefont{Szabo}},
  \bibinfo{author}{\bibfnamefont{S.}~\bibnamefont{Bogar}},
  \bibinfo{author}{\bibfnamefont{F.~J.} \bibnamefont{Nieto}}, \bibnamefont{and}
  \bibinfo{author}{\bibfnamefont{K.~M.} \bibnamefont{Malecki}},
  \bibinfo{journal}{International Journal of Environmental Research and Public
  Health} \textbf{\bibinfo{volume}{11}}, \bibinfo{pages}{3453 }
  (\bibinfo{year}{2014}).

\bibitem[{\citenamefont{Tsai et~al.}(2018)\citenamefont{Tsai, Mchale, Jennings,
  Marquet, Hipp, Leung, and Floyd}}]{Tsai2018}
\bibinfo{author}{\bibfnamefont{W.-L.} \bibnamefont{Tsai}},
  \bibinfo{author}{\bibfnamefont{M.~R.} \bibnamefont{Mchale}},
  \bibinfo{author}{\bibfnamefont{V.}~\bibnamefont{Jennings}},
  \bibinfo{author}{\bibfnamefont{O.}~\bibnamefont{Marquet}},
  \bibinfo{author}{\bibfnamefont{J.~A.} \bibnamefont{Hipp}},
  \bibinfo{author}{\bibfnamefont{Y.-F.} \bibnamefont{Leung}}, \bibnamefont{and}
  \bibinfo{author}{\bibfnamefont{M.~F.} \bibnamefont{Floyd}},
  \bibinfo{journal}{International Journal of Environmental Research and Public
  Health} \textbf{\bibinfo{volume}{15}} (\bibinfo{year}{2018}).

\bibitem[{\citenamefont{Stigsdotter et~al.}(2010)\citenamefont{Stigsdotter,
  Ekholm, Schipperijn, Toftager, Kamper-Jørgensen, and
  Randrup}}]{Stigsdotter2010}
\bibinfo{author}{\bibfnamefont{U.~K.} \bibnamefont{Stigsdotter}},
  \bibinfo{author}{\bibfnamefont{O.}~\bibnamefont{Ekholm}},
  \bibinfo{author}{\bibfnamefont{J.}~\bibnamefont{Schipperijn}},
  \bibinfo{author}{\bibfnamefont{M.}~\bibnamefont{Toftager}},
  \bibinfo{author}{\bibfnamefont{F.}~\bibnamefont{Kamper-Jørgensen}},
  \bibnamefont{and} \bibinfo{author}{\bibfnamefont{T.~B.}
  \bibnamefont{Randrup}}, \bibinfo{journal}{Scandinavian Journal of Public
  Health} \textbf{\bibinfo{volume}{38}}, \bibinfo{pages}{411}
  (\bibinfo{year}{2010}).

\bibitem[{\citenamefont{EOS}(2019)}]{EOS2019NDVI}
\bibinfo{author}{\bibnamefont{EOS}}, \emph{\bibinfo{title}{NDVI FAQ: ALL YOU
  NEED TO KNOW ABOUT NDVI}} (\bibinfo{year}{2019}), \bibinfo{note}{available at
  \url{https://eos.com/blog/ndvi-faq-all-you-need-to-know-about-ndvi/}. Date
  accessed 22 June 2020.}

\bibitem[{\citenamefont{van~den Berg et~al.}(2016)\citenamefont{van~den Berg,
  van Poppel, van Kamp, Andrusaityte, Balseviciene, Cirach, Danileviciute,
  Ellis, Hurst, Masterson et~al.}}]{vandenBerg2016}
\bibinfo{author}{\bibfnamefont{M.}~\bibnamefont{van~den Berg}},
  \bibinfo{author}{\bibfnamefont{M.}~\bibnamefont{van Poppel}},
  \bibinfo{author}{\bibfnamefont{I.}~\bibnamefont{van Kamp}},
  \bibinfo{author}{\bibfnamefont{S.}~\bibnamefont{Andrusaityte}},
  \bibinfo{author}{\bibfnamefont{B.}~\bibnamefont{Balseviciene}},
  \bibinfo{author}{\bibfnamefont{M.}~\bibnamefont{Cirach}},
  \bibinfo{author}{\bibfnamefont{A.}~\bibnamefont{Danileviciute}},
  \bibinfo{author}{\bibfnamefont{N.}~\bibnamefont{Ellis}},
  \bibinfo{author}{\bibfnamefont{G.}~\bibnamefont{Hurst}},
  \bibinfo{author}{\bibfnamefont{D.}~\bibnamefont{Masterson}},
  \bibnamefont{et~al.}, \bibinfo{journal}{Health \& Place}
  \textbf{\bibinfo{volume}{38}}, \bibinfo{pages}{8} (\bibinfo{year}{2016}).

\bibitem[{\citenamefont{Markevych et~al.}(2017)\citenamefont{Markevych,
  Schoierer, Hartig, Chudnovsky, Hystad, Dzhambov, de~Vries, Triguero-Mas,
  Brauer, Nieuwenhuijsen et~al.}}]{Review2017Markevych}
\bibinfo{author}{\bibfnamefont{I.}~\bibnamefont{Markevych}},
  \bibinfo{author}{\bibfnamefont{J.}~\bibnamefont{Schoierer}},
  \bibinfo{author}{\bibfnamefont{T.}~\bibnamefont{Hartig}},
  \bibinfo{author}{\bibfnamefont{A.}~\bibnamefont{Chudnovsky}},
  \bibinfo{author}{\bibfnamefont{P.}~\bibnamefont{Hystad}},
  \bibinfo{author}{\bibfnamefont{A.~M.} \bibnamefont{Dzhambov}},
  \bibinfo{author}{\bibfnamefont{S.}~\bibnamefont{de~Vries}},
  \bibinfo{author}{\bibfnamefont{M.}~\bibnamefont{Triguero-Mas}},
  \bibinfo{author}{\bibfnamefont{M.}~\bibnamefont{Brauer}},
  \bibinfo{author}{\bibfnamefont{M.~J.} \bibnamefont{Nieuwenhuijsen}},
  \bibnamefont{et~al.}, \bibinfo{journal}{Environmental Research}
  \textbf{\bibinfo{volume}{158}}, \bibinfo{pages}{301 } (\bibinfo{year}{2017}).

\bibitem[{\citenamefont{Helliwell et~al.}(2018)\citenamefont{Helliwell, Layard,
  and D.~Sachs}}]{happiness-report2018}
\bibinfo{author}{\bibfnamefont{J.~F.} \bibnamefont{Helliwell}},
  \bibinfo{author}{\bibfnamefont{R.}~\bibnamefont{Layard}}, \bibnamefont{and}
  \bibinfo{author}{\bibfnamefont{J.}~\bibnamefont{D.~Sachs}},
  \bibinfo{journal}{New York: UN Sustainable Development Solutions Network}
  (\bibinfo{year}{2018}).

\bibitem[{\citenamefont{Kahneman and Deaton}(2010)}]{Kahneman2010income}
\bibinfo{author}{\bibfnamefont{D.}~\bibnamefont{Kahneman}} \bibnamefont{and}
  \bibinfo{author}{\bibfnamefont{A.}~\bibnamefont{Deaton}},
  \bibinfo{journal}{Proceedings of the National Academy of Sciences}
  \textbf{\bibinfo{volume}{107}}, \bibinfo{pages}{16489}
  (\bibinfo{year}{2010}).

\bibitem[{\citenamefont{Miura et~al.}(2019)\citenamefont{Miura, Nagai,
  Takeuchi, Ichii, and Yoshioka}}]{Miura2019NDVI}
\bibinfo{author}{\bibfnamefont{T.}~\bibnamefont{Miura}},
  \bibinfo{author}{\bibfnamefont{S.}~\bibnamefont{Nagai}},
  \bibinfo{author}{\bibfnamefont{M.}~\bibnamefont{Takeuchi}},
  \bibinfo{author}{\bibfnamefont{K.}~\bibnamefont{Ichii}}, \bibnamefont{and}
  \bibinfo{author}{\bibfnamefont{H.}~\bibnamefont{Yoshioka}},
  \bibinfo{journal}{Scientific Reports} \textbf{\bibinfo{volume}{9}}
  (\bibinfo{year}{2019}).

\bibitem[{\citenamefont{IMF}(2018)}]{IMF2018}
\bibinfo{author}{\bibnamefont{IMF}}, \emph{\bibinfo{title}{World Economic
  Outlook Database}} (\bibinfo{year}{2018}), \bibinfo{note}{available at
  \url{https://www.imf.org/en/Publications/SPROLLS/world-economic-outlook-databases}.
  Date accessed 16 November 2020}.

\bibitem[{\citenamefont{Easterlin et~al.}(2010)\citenamefont{Easterlin, McVey,
  Switek, Sawangfa, and Zweig}}]{Easterlin2010}
\bibinfo{author}{\bibfnamefont{R.~A.} \bibnamefont{Easterlin}},
  \bibinfo{author}{\bibfnamefont{L.~A.} \bibnamefont{McVey}},
  \bibinfo{author}{\bibfnamefont{M.}~\bibnamefont{Switek}},
  \bibinfo{author}{\bibfnamefont{O.}~\bibnamefont{Sawangfa}}, \bibnamefont{and}
  \bibinfo{author}{\bibfnamefont{J.~S.} \bibnamefont{Zweig}},
  \bibinfo{journal}{Proceedings of the National Academy of Sciences}
  \textbf{\bibinfo{volume}{107}}, \bibinfo{pages}{22463}
  (\bibinfo{year}{2010}).

\bibitem[{\citenamefont{Preacher et~al.}(2007)\citenamefont{Preacher, Rucker,
  and Hayes}}]{Preacher2007}
\bibinfo{author}{\bibfnamefont{K.~J.} \bibnamefont{Preacher}},
  \bibinfo{author}{\bibfnamefont{D.~D.} \bibnamefont{Rucker}},
  \bibnamefont{and} \bibinfo{author}{\bibfnamefont{A.~F.} \bibnamefont{Hayes}},
  \bibinfo{journal}{Multivariate Behavioral Research}
  \textbf{\bibinfo{volume}{42}}, \bibinfo{pages}{185–227}
  (\bibinfo{year}{2007}).

\bibitem[{\citenamefont{Foley and Kistemann}(2015)}]{Foley2015}
\bibinfo{author}{\bibfnamefont{R.}~\bibnamefont{Foley}} \bibnamefont{and}
  \bibinfo{author}{\bibfnamefont{T.}~\bibnamefont{Kistemann}},
  \bibinfo{journal}{Health \& Place} \textbf{\bibinfo{volume}{35}},
  \bibinfo{pages}{157–165} (\bibinfo{year}{2015}).

\bibitem[{\citenamefont{Raymond et~al.}(2016)\citenamefont{Raymond, Gottwald,
  and Kytta}}]{Raymond2016}
\bibinfo{author}{\bibfnamefont{C.~M.} \bibnamefont{Raymond}},
  \bibinfo{author}{\bibfnamefont{S.}~\bibnamefont{Gottwald}}, \bibnamefont{and}
  \bibinfo{author}{\bibfnamefont{M.}~\bibnamefont{Kytta}},
  \bibinfo{journal}{Landscape and Urban Planning}
  \textbf{\bibinfo{volume}{153}}, \bibinfo{pages}{198–208}
  (\bibinfo{year}{2016}).

\bibitem[{\citenamefont{Groff and McCord}(2012)}]{groff2012role}
\bibinfo{author}{\bibfnamefont{E.}~\bibnamefont{Groff}} \bibnamefont{and}
  \bibinfo{author}{\bibfnamefont{E.~S.} \bibnamefont{McCord}},
  \bibinfo{journal}{Security journal} \textbf{\bibinfo{volume}{25}},
  \bibinfo{pages}{1} (\bibinfo{year}{2012}).

\bibitem[{\citenamefont{Han et~al.}(2018)\citenamefont{Han, Cohen, Derose, Li,
  and Williamson}}]{han2018violent}
\bibinfo{author}{\bibfnamefont{B.}~\bibnamefont{Han}},
  \bibinfo{author}{\bibfnamefont{D.~A.} \bibnamefont{Cohen}},
  \bibinfo{author}{\bibfnamefont{K.~P.} \bibnamefont{Derose}},
  \bibinfo{author}{\bibfnamefont{J.}~\bibnamefont{Li}}, \bibnamefont{and}
  \bibinfo{author}{\bibfnamefont{S.}~\bibnamefont{Williamson}},
  \bibinfo{journal}{American journal of preventive medicine}
  \textbf{\bibinfo{volume}{54}}, \bibinfo{pages}{352} (\bibinfo{year}{2018}).

\bibitem[{\citenamefont{Ugolini et~al.}(2020)\citenamefont{Ugolini, Massetti,
  Calaza-Martínez, Cariñanos, Dobbs, Ostoić, Marin, Pearlmutter, Saaroni,
  Šaulienė et~al.}}]{Ugolini2020}
\bibinfo{author}{\bibfnamefont{F.}~\bibnamefont{Ugolini}},
  \bibinfo{author}{\bibfnamefont{L.}~\bibnamefont{Massetti}},
  \bibinfo{author}{\bibfnamefont{P.}~\bibnamefont{Calaza-Martínez}},
  \bibinfo{author}{\bibfnamefont{P.}~\bibnamefont{Cariñanos}},
  \bibinfo{author}{\bibfnamefont{C.}~\bibnamefont{Dobbs}},
  \bibinfo{author}{\bibfnamefont{S.~K.} \bibnamefont{Ostoić}},
  \bibinfo{author}{\bibfnamefont{A.~M.} \bibnamefont{Marin}},
  \bibinfo{author}{\bibfnamefont{D.}~\bibnamefont{Pearlmutter}},
  \bibinfo{author}{\bibfnamefont{H.}~\bibnamefont{Saaroni}},
  \bibinfo{author}{\bibfnamefont{I.}~\bibnamefont{Šaulienė}},
  \bibnamefont{et~al.}, \bibinfo{journal}{Urban Forestry \& Urban Greening}
  \textbf{\bibinfo{volume}{56}}, \bibinfo{pages}{126888}
  (\bibinfo{year}{2020}).

\bibitem[{\citenamefont{Lolli et~al.}(2020)\citenamefont{Lolli, Chen, Wang, and
  Vivone}}]{Lolli2020}
\bibinfo{author}{\bibfnamefont{S.}~\bibnamefont{Lolli}},
  \bibinfo{author}{\bibfnamefont{Y.-C.} \bibnamefont{Chen}},
  \bibinfo{author}{\bibfnamefont{S.-H.} \bibnamefont{Wang}}, \bibnamefont{and}
  \bibinfo{author}{\bibfnamefont{G.}~\bibnamefont{Vivone}},
  \bibinfo{journal}{Scientific Reports} \textbf{\bibinfo{volume}{10}},
  \bibinfo{pages}{16213} (\bibinfo{year}{2020}).

\bibitem[{\citenamefont{Hedblom et~al.}(2019)\citenamefont{Hedblom, Gunnarsson,
  Iravani, Knez, Schaefer, Thorsson, and Lundstr\"{o}m}}]{Hedblom2019}
\bibinfo{author}{\bibfnamefont{M.}~\bibnamefont{Hedblom}},
  \bibinfo{author}{\bibfnamefont{B.}~\bibnamefont{Gunnarsson}},
  \bibinfo{author}{\bibfnamefont{B.}~\bibnamefont{Iravani}},
  \bibinfo{author}{\bibfnamefont{I.}~\bibnamefont{Knez}},
  \bibinfo{author}{\bibfnamefont{M.}~\bibnamefont{Schaefer}},
  \bibinfo{author}{\bibfnamefont{P.}~\bibnamefont{Thorsson}}, \bibnamefont{and}
  \bibinfo{author}{\bibfnamefont{J.~N.} \bibnamefont{Lundstr\"{o}m}},
  \bibinfo{journal}{Scientific Reports} \textbf{\bibinfo{volume}{9}},
  \bibinfo{pages}{10113} (\bibinfo{year}{2019}).

\bibitem[{\citenamefont{Ewing}(2008)}]{Ewing2008sprawl}
\bibinfo{author}{\bibfnamefont{R.~H.} \bibnamefont{Ewing}},
  \emph{\bibinfo{title}{Characteristics, Causes, and Effects of Sprawl: A
  Literature Review}} (\bibinfo{publisher}{Springer US},
  \bibinfo{address}{Boston, MA}, \bibinfo{year}{2008}).

\bibitem[{\citenamefont{Liu et~al.}(2020)\citenamefont{Liu, Huang, Xu, Li, Li,
  Ciais, Lin, Gong, Ziegler, Chen et~al.}}]{Liu2020greenrecovery}
\bibinfo{author}{\bibfnamefont{X.}~\bibnamefont{Liu}},
  \bibinfo{author}{\bibfnamefont{Y.}~\bibnamefont{Huang}},
  \bibinfo{author}{\bibfnamefont{X.}~\bibnamefont{Xu}},
  \bibinfo{author}{\bibfnamefont{X.}~\bibnamefont{Li}},
  \bibinfo{author}{\bibfnamefont{X.}~\bibnamefont{Li}},
  \bibinfo{author}{\bibfnamefont{P.}~\bibnamefont{Ciais}},
  \bibinfo{author}{\bibfnamefont{P.}~\bibnamefont{Lin}},
  \bibinfo{author}{\bibfnamefont{K.}~\bibnamefont{Gong}},
  \bibinfo{author}{\bibfnamefont{A.~D.} \bibnamefont{Ziegler}},
  \bibinfo{author}{\bibfnamefont{A.}~\bibnamefont{Chen}}, \bibnamefont{et~al.},
  \bibinfo{journal}{Nature Sustainability} \textbf{\bibinfo{volume}{3}},
  \bibinfo{pages}{564} (\bibinfo{year}{2020}).

\bibitem[{\citenamefont{Allen et~al.}(2010)\citenamefont{Allen, Macalady,
  Chenchouni, Bachelet, McDowell, Vennetier, Kitzberger, Rigling, Breshears,
  Hogg et~al.}}]{Allen2010climatechange}
\bibinfo{author}{\bibfnamefont{C.~D.} \bibnamefont{Allen}},
  \bibinfo{author}{\bibfnamefont{A.~K.} \bibnamefont{Macalady}},
  \bibinfo{author}{\bibfnamefont{H.}~\bibnamefont{Chenchouni}},
  \bibinfo{author}{\bibfnamefont{D.}~\bibnamefont{Bachelet}},
  \bibinfo{author}{\bibfnamefont{N.}~\bibnamefont{McDowell}},
  \bibinfo{author}{\bibfnamefont{M.}~\bibnamefont{Vennetier}},
  \bibinfo{author}{\bibfnamefont{T.}~\bibnamefont{Kitzberger}},
  \bibinfo{author}{\bibfnamefont{A.}~\bibnamefont{Rigling}},
  \bibinfo{author}{\bibfnamefont{D.~D.} \bibnamefont{Breshears}},
  \bibinfo{author}{\bibfnamefont{E.~H.} \bibnamefont{Hogg}},
  \bibnamefont{et~al.}, \bibinfo{journal}{Forest Ecology and Management}
  \textbf{\bibinfo{volume}{259}}, \bibinfo{pages}{660} (\bibinfo{year}{2010}).

\bibitem[{\citenamefont{Pretzsch et~al.}(2017)\citenamefont{Pretzsch, Biber,
  Uhl, Dahlhausen, Sch\"{u}tze, Perkins, R\"{o}tzer, Caldentey, Koike, van Con
  et~al.}}]{Pretzsch2017climatechange}
\bibinfo{author}{\bibfnamefont{H.}~\bibnamefont{Pretzsch}},
  \bibinfo{author}{\bibfnamefont{P.}~\bibnamefont{Biber}},
  \bibinfo{author}{\bibfnamefont{E.}~\bibnamefont{Uhl}},
  \bibinfo{author}{\bibfnamefont{J.}~\bibnamefont{Dahlhausen}},
  \bibinfo{author}{\bibfnamefont{G.}~\bibnamefont{Sch\"{u}tze}},
  \bibinfo{author}{\bibfnamefont{D.}~\bibnamefont{Perkins}},
  \bibinfo{author}{\bibfnamefont{T.}~\bibnamefont{R\"{o}tzer}},
  \bibinfo{author}{\bibfnamefont{J.}~\bibnamefont{Caldentey}},
  \bibinfo{author}{\bibfnamefont{T.}~\bibnamefont{Koike}},
  \bibinfo{author}{\bibfnamefont{T.}~\bibnamefont{van Con}},
  \bibnamefont{et~al.}, \bibinfo{journal}{Scientific Reports}
  \textbf{\bibinfo{volume}{7}} (\bibinfo{year}{2017}).

\bibitem[{\citenamefont{Nowak et~al.}(2013)\citenamefont{Nowak, Greenfield,
  Hoehn, and Lapoint}}]{Nowak2013carbon}
\bibinfo{author}{\bibfnamefont{D.~J.} \bibnamefont{Nowak}},
  \bibinfo{author}{\bibfnamefont{E.~J.} \bibnamefont{Greenfield}},
  \bibinfo{author}{\bibfnamefont{R.~E.} \bibnamefont{Hoehn}}, \bibnamefont{and}
  \bibinfo{author}{\bibfnamefont{E.}~\bibnamefont{Lapoint}},
  \bibinfo{journal}{Environmental Pollution} \textbf{\bibinfo{volume}{178}},
  \bibinfo{pages}{229} (\bibinfo{year}{2013}).

\bibitem[{\citenamefont{Houlden et~al.}(2017)\citenamefont{Houlden, Weich, and
  Jarvis}}]{houlden2017cross}
\bibinfo{author}{\bibfnamefont{V.}~\bibnamefont{Houlden}},
  \bibinfo{author}{\bibfnamefont{S.}~\bibnamefont{Weich}}, \bibnamefont{and}
  \bibinfo{author}{\bibfnamefont{S.}~\bibnamefont{Jarvis}},
  \bibinfo{journal}{BMC public health} \textbf{\bibinfo{volume}{17}},
  \bibinfo{pages}{460} (\bibinfo{year}{2017}).

\bibitem[{\citenamefont{Seiferling et~al.}(2017)\citenamefont{Seiferling, Naik,
  Ratti, and Proulx}}]{Seiferling2017}
\bibinfo{author}{\bibfnamefont{I.}~\bibnamefont{Seiferling}},
  \bibinfo{author}{\bibfnamefont{N.}~\bibnamefont{Naik}},
  \bibinfo{author}{\bibfnamefont{C.}~\bibnamefont{Ratti}}, \bibnamefont{and}
  \bibinfo{author}{\bibfnamefont{R.}~\bibnamefont{Proulx}},
  \bibinfo{journal}{Landscape and Urban Planning}
  \textbf{\bibinfo{volume}{165}}, \bibinfo{pages}{93 } (\bibinfo{year}{2017}).

\bibitem[{\citenamefont{Huete}(1988)}]{Huete1988SAVI}
\bibinfo{author}{\bibfnamefont{A.~R.} \bibnamefont{Huete}},
  \bibinfo{journal}{Remote Sensing of Environment}
  \textbf{\bibinfo{volume}{25}}, \bibinfo{pages}{295 } (\bibinfo{year}{1988}).

\bibitem[{\citenamefont{Jiang et~al.}(2008)\citenamefont{Jiang, Huete, Didan,
  and Miura}}]{Jiang2008EVI2}
\bibinfo{author}{\bibfnamefont{Z.}~\bibnamefont{Jiang}},
  \bibinfo{author}{\bibfnamefont{A.~R.} \bibnamefont{Huete}},
  \bibinfo{author}{\bibfnamefont{K.}~\bibnamefont{Didan}}, \bibnamefont{and}
  \bibinfo{author}{\bibfnamefont{T.}~\bibnamefont{Miura}},
  \bibinfo{journal}{Remote Sensing of Environment}
  \textbf{\bibinfo{volume}{112}}, \bibinfo{pages}{3833 }
  (\bibinfo{year}{2008}).

\bibitem[{\citenamefont{Buchhorn et~al.}(2020)\citenamefont{Buchhorn, Smets,
  Bertels, De~Roo, Lesiv, Tsendbazar, Herold, and Fritz}}]{Copernicus2020}
\bibinfo{author}{\bibfnamefont{M.}~\bibnamefont{Buchhorn}},
  \bibinfo{author}{\bibfnamefont{B.}~\bibnamefont{Smets}},
  \bibinfo{author}{\bibfnamefont{L.}~\bibnamefont{Bertels}},
  \bibinfo{author}{\bibfnamefont{B.}~\bibnamefont{De~Roo}},
  \bibinfo{author}{\bibfnamefont{M.}~\bibnamefont{Lesiv}},
  \bibinfo{author}{\bibfnamefont{N.-E.} \bibnamefont{Tsendbazar}},
  \bibinfo{author}{\bibfnamefont{M.}~\bibnamefont{Herold}}, \bibnamefont{and}
  \bibinfo{author}{\bibfnamefont{S.}~\bibnamefont{Fritz}},
  \emph{\bibinfo{title}{Copernicus Global Land Service: Land Cover 100m:
  collection 3 epoch 2015, Globe}} (\bibinfo{year}{2020}),
  \bibinfo{note}{available at \url{https://lcviewer.vito.be/download}}.

\end{thebibliography}

\end{document}


\maketitle

\normalsize

\newpage

\section{Data}\label{sec:data}

\subsection{Data description}

Table \ref{tab:si_dataset1} and \ref{tab:si_dataset2} describe the dataset used in this study. The happiness scores were obtained from the World Happiness Report, which was averaged over three years to adjust for short-term fluctuations. The average happiness score is 6.373, with a maximum of 7.769 for Finland and a minimum of 4.549 for Iran. UGS is calculated from Sentinel-2 satellite imagery data, and GDP per capita (PPP) data is obtained from the IMF estimation.

In this research, we used the data of 60 developed countries selected by comparing the HDI of the countries. Andorra, Bahamas, Barbados, Brunei, Cyprus, Lichtenstein, Palau, and Seychelles are excluded from the analysis due to a lack of data for happiness. 

\begin{longtable}{r|cc|ccc}
\hline
                 Country &  City counts & Population [\%] &  Happiness &    UGS &      log-GDP \\
\hline
              Finland &          1 &     28.23 &          7.77 &    5.73 &  10.70 \\
              Iceland &          1 &     38.05 &          7.49 &    5.47 &  10.87 \\
            Lithuania &          1 &     19.25 &          6.15 &    5.46 &  10.44 \\
          New Zealand &          1 &     34.57 &          7.31 &    5.33 &  10.60 \\
             Slovenia &          1 &     13.99 &          6.12 &    5.32 &  10.45 \\
              Croatia &          1 &     19.82 &          5.43 &    5.23 &  10.11 \\
           Montenegro &          1 &     31.07 &          5.52 &    5.21 &   9.85 \\
                Italy &          1 &      7.21 &          6.22 &    5.17 &  10.56 \\
             Slovakia &          2 &     12.25 &          6.20 &    5.16 &  10.46 \\
              Estonia &          1 &     33.12 &          5.89 &    5.15 &  10.39 \\
        United States &          3 &     12.76 &          6.89 &    5.13 &  11.03 \\
               Latvia &          1 &     32.95 &          5.94 &    5.05 &  10.28 \\
               Sweden &          2 &     15.01 &          7.34 &    5.00 &  10.88 \\
          Switzerland &          4 &     10.87 &          7.48 &    4.98 &  11.04 \\
               Norway &          1 &     12.80 &          7.54 &    4.97 &  11.19 \\
               Canada &          1 &     18.26 &          7.28 &    4.96 &  10.80 \\
               Serbia &          1 &     23.99 &          5.60 &    4.93 &   9.67 \\
               Poland &          4 &     10.09 &          6.16 &    4.88 &  10.34 \\
              Germany &          5 &     10.54 &          7.02 &    4.79 &  10.85 \\
              Hungary &          1 &     17.89 &          5.82 &    4.78 &  10.31 \\
       Czech Republic &          1 &     12.13 &          6.85 &    4.75 &  10.50 \\
             Portugal &          3 &     11.67 &          5.69 &    4.72 &  10.32 \\
             Bulgaria &          1 &     18.69 &          5.01 &    4.70 &  10.02 \\
            Australia &          1 &     19.55 &          7.23 &    4.69 &  10.86 \\
          Netherlands &          3 &     10.71 &          7.49 &    4.52 &  10.91 \\
           Luxembourg &          1 &     30.40 &          7.09 &    4.49 &  11.59 \\
              Ireland &          1 &     11.62 &          7.02 &    4.36 &  11.24 \\
       United Kingdom &          1 &     13.42 &          7.05 &    4.28 &  10.72 \\
  Trinidad and Tobago &          1 &     12.76 &          6.19 &    4.25 &  10.46 \\
              Uruguay &          1 &     39.42 &          6.29 &    4.16 &  10.06 \\
\hline
\caption{\label{tab:si_dataset1}Data used in the study. Countries are ordered by UGS. We aggregate city-level data to cover at least 10\% of total population.}
\end{longtable}

\newpage

\begin{longtable}{r|cc|ccc}
\hline
                 Country &  City counts & Population [\%] &  Happiness &    UGS &      log-GDP \\
\hline
                Spain &          1 &     13.97 &          6.35 &    4.15 &  10.59 \\
               Russia &          2 &     12.50 &          5.65 &    4.12 &  10.24 \\
              Belarus &          1 &     20.82 &          5.32 &    4.12 &   9.82 \\
              Austria &          1 &     21.42 &          7.25 &    4.11 &  10.83 \\
               Panama &          1 &     27.74 &          6.32 &    4.06 &  10.15 \\
           Kazakhstan &          1 &     11.13 &          5.81 &    4.06 &  10.19 \\
              Albania &          1 &     23.32 &          4.72 &    4.06 &   9.51 \\
            Mauritius &          1 &     29.03 &          5.89 &    3.94 &  10.05 \\
           Costa Rica &          1 &     32.68 &          7.17 &    3.93 &   9.79 \\
              Belgium &          1 &     10.58 &          6.92 &    3.91 &  10.76 \\
              Denmark &          1 &     10.78 &          7.60 &    3.89 &  10.82 \\
              Romania &          2 &     10.93 &          6.07 &    3.87 &  10.13 \\
               France &          1 &     10.62 &          6.59 &    3.72 &  10.72 \\
             Malaysia &          1 &     12.19 &          5.34 &    3.64 &  10.31 \\
            Argentina &          2 &     10.13 &          6.09 &    3.33 &   9.98 \\
               Turkey &          1 &     18.34 &          5.37 &    3.28 &  10.05 \\
               Greece &          1 &     24.11 &          5.29 &    3.28 &  10.30 \\
                Malta &          3 &     13.05 &          6.73 &    3.17 &  10.64 \\
                Chile &          1 &     30.54 &          6.45 &    3.05 &  10.15 \\
                Japan &          1 &     10.63 &          5.89 &    3.03 &  10.63 \\
                 Iran &          1 &     10.86 &          4.55 &    2.90 &   9.91 \\
            Singapore &          1 &    100.00 &          6.26 &    2.87 &  11.45 \\
          South Korea &          1 &     19.00 &          5.89 &    2.70 &  10.64 \\
               Israel &          1 &     10.75 &          7.14 &    2.65 &  10.53 \\
 United Arab Emirates &          1 &     35.82 &          6.82 &    2.23 &  11.17 \\
         Saudi Arabia &          1 &     19.49 &          6.37 &    2.06 &  10.95 \\
                 Oman &          1 &     32.66 &          6.85 &    2.05 &  10.73 \\
               Kuwait &          1 &     12.79 &          6.02 &    1.91 &  11.22 \\
                Qatar &          1 &     39.77 &          6.37 &    1.23 &  11.82 \\
              Bahrain &          1 &     38.47 &          6.20 &    0.54 &  10.86 \\
\hline
\caption{\label{tab:si_dataset2}Data used in the study. Countries are ordered by UGS. We aggregate city-level data to cover at least 10\% of total population.}
\end{longtable}

\section{Robustness of the regression}
The result of the regression is robust for any green space measure. In table \ref{tab:si_robust}, all nine green space measures explain happiness along with GDP, while logarithmic NDVI per capita in the model (5) displays the most considerable value of adjusted $R^2$ compared to other models.

\begin{landscape}
\begin{table}[ht]
\centering
\begin{tabular}{c|c c c c c c c c c}
\hline
	&	(1)	&	(2)	&	(3)	&	(4)	&	(5)	&	(6)	&	(7)	&	(8)	&	(9)	\\
\hline
\hline
GDPLN	&	1.0684\textsuperscript{***}	&	1.0198\textsuperscript{***}	&	1.1118\textsuperscript{***}	&	1.0333\textsuperscript{***}	&	1.1319\textsuperscript{***}	&	1.1138\textsuperscript{***}	&	1.0347\textsuperscript{***}	&	1.1001\textsuperscript{***}	&	1.0292\textsuperscript{***}	\\
	&	(0.6455)	&	(0.6291)	&	(0.6507)	&	(0.6264)	&	(0.6234)	&	(0.6501)	&	(0.6266)	&	(0.6453)	&	(0.6256)	\\
GreenRatio	&	0.0101\textsuperscript{**}	&	-	&	-	&	-	&	-	&	-	&	-	&	-	&	-	\\
	&	(0.0177)	&		&		&		&		&		&		&		&		\\
GreenperCapita	&	-	&	0.0020\textsuperscript{**}	&	-	&	-	&	-	&	-	&	-	&	-	&	-	\\
	&		&	(0.0031)	&		&		&		&		&		&		&		\\
NDVIMean	&	-	&	-	&	1.8166\textsuperscript{**}	&	-	&	-	&	-	&	-	&	-	&	-	\\
	&		&		&	(2.8613)	&		&		&		&		&		&		\\
NDVIperCapita	&	-	&	-	&	-	&	0.0029\textsuperscript{***}	&	-	&	-	&	-	&	-	&	-	\\
	&		&		&		&	(0.0042)	&		&		&		&		&		\\
NDVILN	&	-	&	-	&	-	&	-	&	0.2249\textsuperscript{***}	&	-	&	-	&	-	&	-	\\
	&		&		&		&		&	(0.2643)	&		&		&		&		\\
SAVIMean	&	-	&	-	&	-	&	-	&	-	&	1.2267\textsuperscript{**}	&	-	&	-	&	-	\\
	&		&		&		&		&		&	(1.9051)	&		&		&		\\
SAVIperCapita	&	-	&	-	&	-	&	-	&	-	&	-	&	0.0019\textsuperscript{***}	&	-	&	-	\\
	&		&		&		&		&		&		&	(0.0028)	&		&		\\
EVI2Mean	&	-	&	-	&	-	&	-	&	-	&	-	&	-	&	0.9462\textsuperscript{**}	&	-	\\
	&		&		&		&		&		&		&		&	(1.4709)	&		\\
EVI2perCapita	&	-	&	-	&	-	&	-	&	-	&	-	&	-	&	-	&	0.0016\textsuperscript{***}	\\
	&		&		&		&		&		&		&		&		&	(0.0022)	\\
Const	&	-5.1875***	&	-4.5886***	&	-5.8446***	&	-4.7804***	&	-6.4709***	&	-5.8728***	&	-4.7950***	&	-5.6947***	&	-4.7325***	\\
	&	(6.9109)	&	(6.6518)	&	(7.0963)	&	(6.6389)	&	(6.8998)	&	(7.0905)	&	(6.6423)	&	(7.0009)	&	(6.6259)	\\
\hline
Adjusted $R^2$	&	0.4248	&	0.4407	&	0.4363	&	0.4466	&	0.4786	&	0.4379	&	0.4465	&	0.4378	&	0.4476	\\
Observations & 60 & 60 & 60 & 60 & 60 & 60 & 60 & 60 & 60\\
\hline
\end{tabular} 
\caption{\label{tab:si_robust}Regression analysis of happiness with different green space measures, \textsuperscript{***}p<0.01; \textsuperscript{**}p<0.05; \textsuperscript{*}p<0.1.}
\end{table}
\end{landscape}

\section{Regional influence}

\begin{figure}[ht]
\centering
\includegraphics[width=\linewidth]{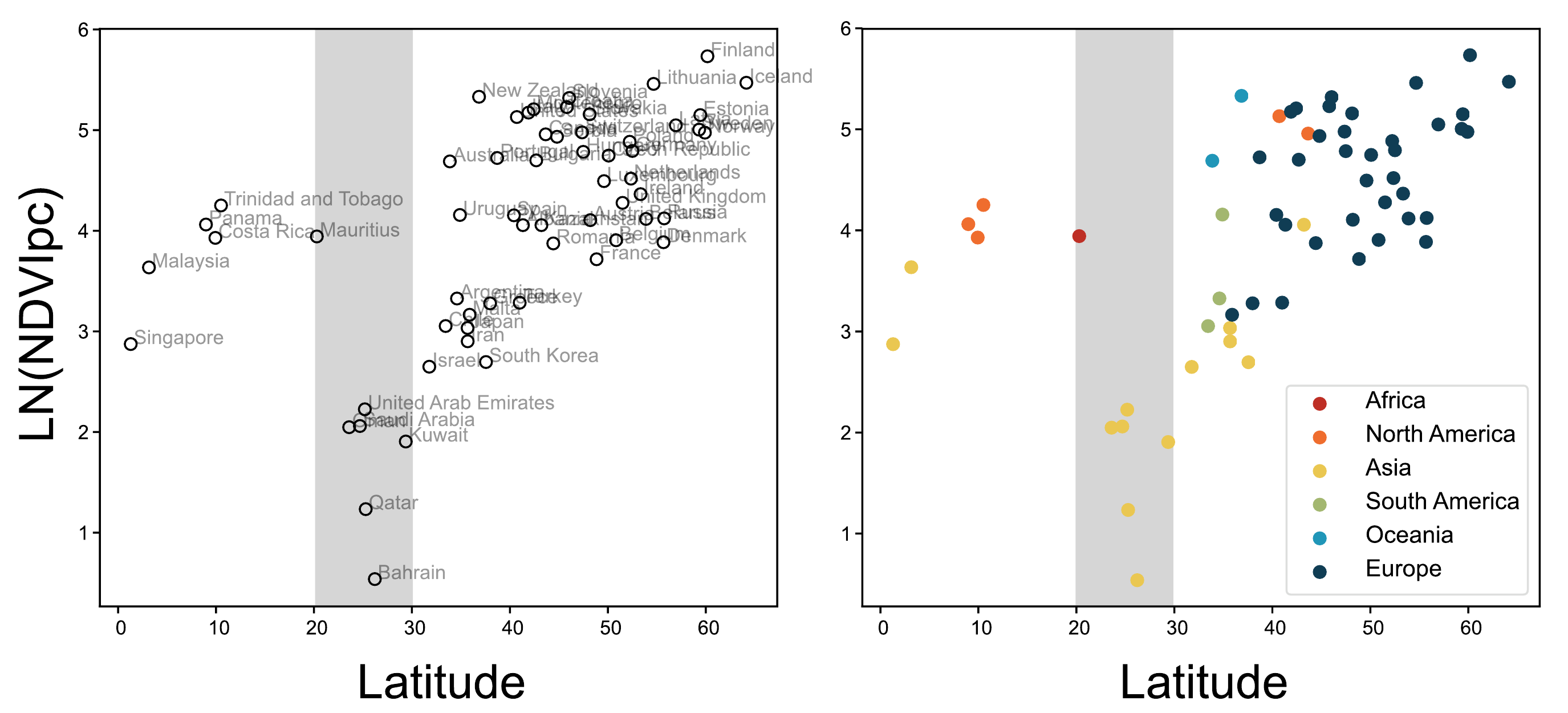}
\caption{Scatter plot of UGS and latitude with country (left) and continent (right) marked. Gray area represent the dry climate region.}
\label{fig:si_region}
\end{figure}

Regional characteristics affect the level of green space. Figure~\ref{fig:si_region} describes the change of USG by latitude. Countries with a tropical climate such as Southeastern Asia, the Caribbean, and Eastern Africa show a relatively high UGS score. In contrast, Western Asian countries show a relatively low UGS score since they are in a dry climate. The UGS score further increases in higher latitudes.

In Table~\ref{tab:si_region}, model(3) includes the latitude of the most populated city, model (4-5) includes dummy variables that tell whether the countries in Western Asia or the dry climate region. These models show that including regional factors does not improve the model.

\begin{table}[ht]
\centering
\small
\begin{tabular}{c|c c c c c}
\hline
	&	(1)	&	(2)	&	(3) &   (4) &   (5)\\
\hline
\hline
GDP	&	1.0120\textsuperscript{***}	&	1.1319\textsuperscript{***}	&	1.1275\textsuperscript{***}	&	1.1142\textsuperscript{***}	&	1.1347\textsuperscript{***}	\\
	&	(0.6603)	&	(0.6234)	&	(0.6413)	&	(0.6433)	&	(0.6297)	\\
UGS	&	-	&	0.2249\textsuperscript{***}	&	0.2181\textsuperscript{**}	&	0.2585\textsuperscript{***}	&	0.2055\textsuperscript{**}	\\
	&		&	(0.2643)	&	(0.3302)	&	(0.3770)	&	(0.3641)	\\
Latitude	&	-	&	-	&	0.0009	&	-	&	-	\\
	&		&		&	(0.0250)	&		&		\\
Western Asia	&	-	&	-	&	-	&	0.1595	&	-	\\
	&		&		&		&	(1.2679)	&		\\
Dry Climate	&	-	&	-	&	-	&	-	&	-0.0885	\\
	&		&		&		&		& (1.1301)		\\
Const	&	-4.2945\textsuperscript{**}	&	-6.4709\textsuperscript{***}	&	-6.4326\textsuperscript{***}	&	-6.4422\textsuperscript{***}	&	-6.4081\textsuperscript{***}	\\
	&	(6.9672)	&	(6.8998)	&	(7.0474)	&	(6.9518)	&	(7.0037)	\\
\hline
Adjusted $R^2$	&	0.3832	&	0.4786	&	0.4695	&	0.4717	&	0.4702	\\
Observations & 60 & 60 & 60 & 60 & 60\\
\hline
\end{tabular}
\caption{\label{tab:si_region}Regression analysis of happiness with region variables. \textsuperscript{***}p<0.01; \textsuperscript{**}p<0.05; \textsuperscript{*}p<0.1.}
\end{table}

\newpage

\section{Distribution of green space}

\begin{figure}[ht]
\centering
\includegraphics[width=\linewidth]{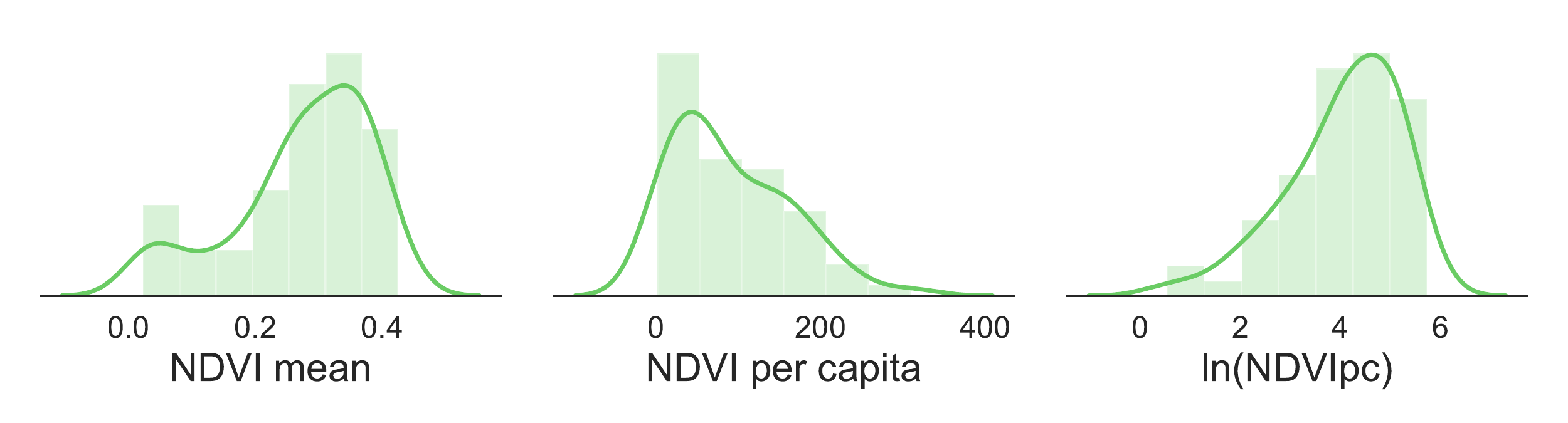}
\caption{Distribution plot of NDVI mean, NDVI per capita, logarithmic NDVI per capita}
\label{fig:si_dist}
\end{figure}

Figure~\ref{fig:si_dist} describes the distribution of three green space measures. NDVIavg (average NDVI) is calculated by taking the mean NDVI values over the built-up area, representing how much greenery cities have. NDVIpc (average NDVI per capita) is obtained by dividing the total NDVI by the total population. NDVI per capita describes how much green space is provided to a population. However, NDVI per capita shows a skewed distribution, which is not appropriate for regression analysis. Therefore, we log on to NDVI per capita to get a normal-like distribution of the green space measure.

\section{Residual analysis}

We perform a residual analysis of the regression model in Table 1 to check whether the model is reasonable. First, we need to check the autocorrelation of the residuals by using Durbin-Watson statistics. The Durbin-Watson statistics show a value of 1.918, which indicates there are no autocorrelations between the residuals. Second, we check for the normality of the residuals. The distribution and Q-Q plot of the residuals shows that the residuals satisfy the normality condition. Finally, we check for the equality of variance by finding outliers using Cook's distance. The figure shows that every point has a value of less than 1, indicating acceptable values.

\begin{figure}[ht]
\centering
\includegraphics[width=\linewidth]{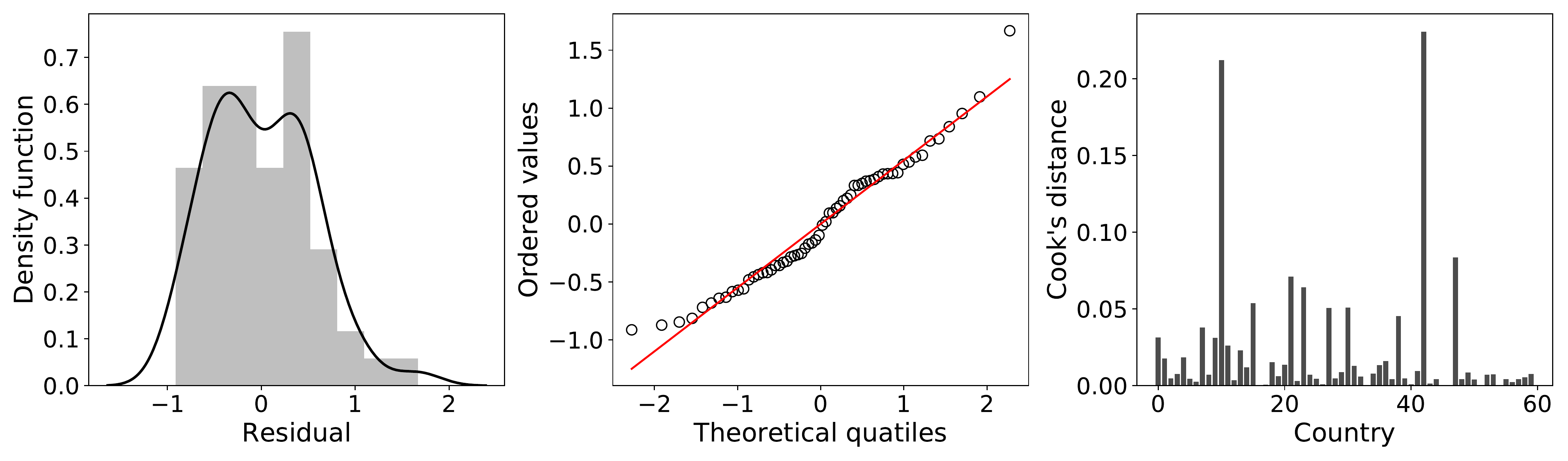}
\caption{Residual analysis of the regression model. (left) The distribution, (middle) Q-Q plot, and (right) cook's distance of residuals.}
\label{fig:si_residual}
\end{figure}

\newpage

\section{The effect of GDP on green-happiness relation}

\begin{figure}[ht]
\centering
\includegraphics[width=0.6\linewidth]{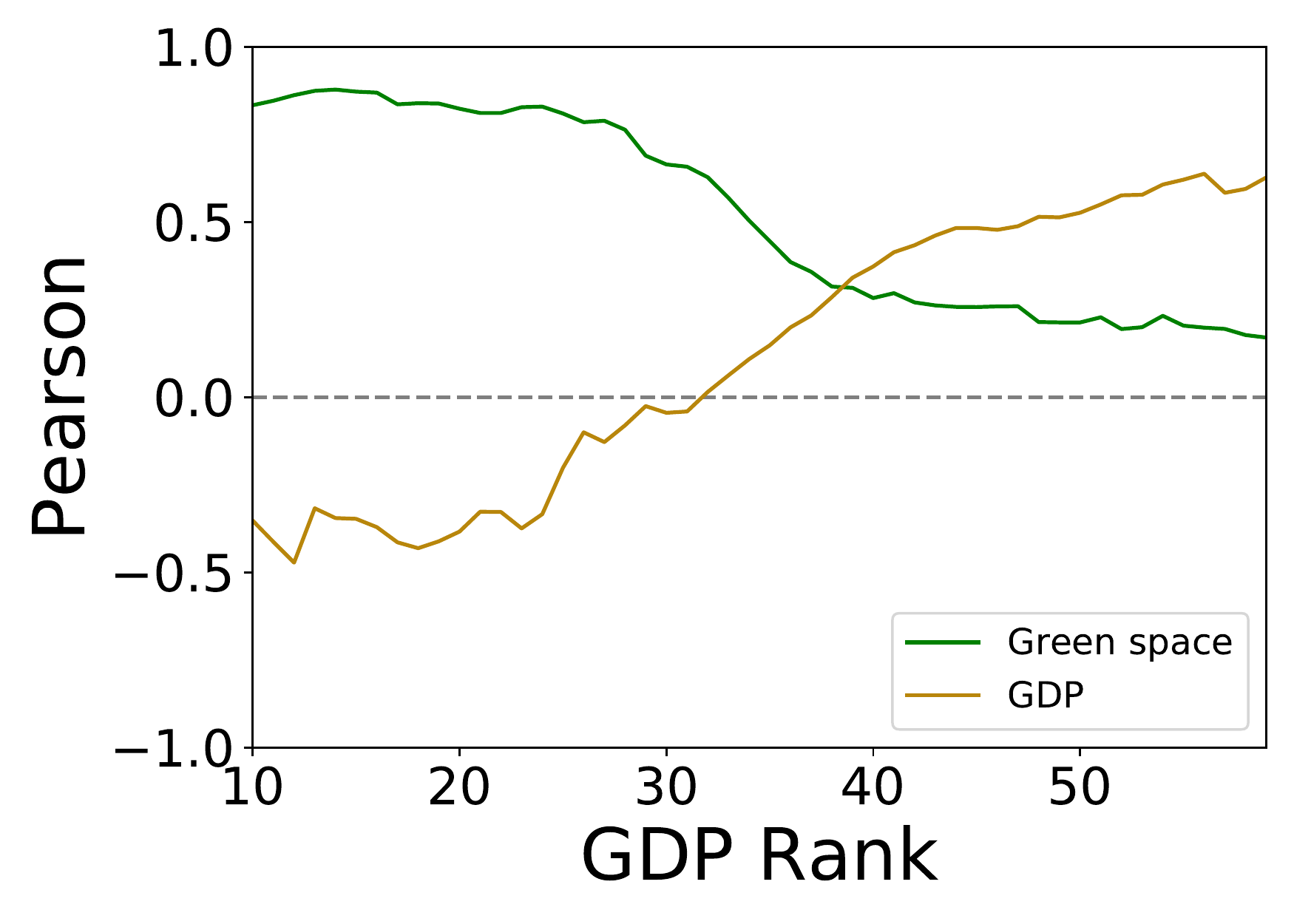}
\caption{Changes of the Pearson correlation between urban green space and happiness for different sets of GDP rank with increasing window size from top 10 to 60.}
\label{fig:si_corr}
\end{figure}

We can check for a similar result of Fig. 3(c) in the manuscript by calculating the Pearson correlation instead of the regression coefficient. Figure \ref{fig:si_corr} shows a similar diminishing effect of green space as the group contains lower GDP countries. In contrast, log-GDP shows the most strong correlations for the entire dataset containing lower GDP groups.

\section{Happiness Report variables}

World Happiness Report describes happiness with six main variables: \textit{GDP, social support, life expectancy, freedom, generosity}, and \textit{corruption perceptions}. \textit{Social support} and \textit{freedom} are based on binary responses (yes or no) to World Gallup Poll (WGP) questions; ``If you were in trouble, do you have relatives or friends you can count on to help you whenever you need them, or not?'', and ``Are you satisfied or dissatisfied with your freedom to choose what you do with your life?'', respectively. \textit{generosity} is the residual of regression for responses for a WGP question ``Have you donated money to a charity in the past month?” on GDP per capita. \textit{Corruption perceptions} is based on the response to WGP question, ``Is corruption widespread throughout the government or not?'' and ``Is corruption widespread within businesses or not?'' \textit{Life expectancy} is based on the Global Health Observatory data from World Health Organization (WHO).

\begin{figure}[ht]
\centering
\includegraphics[width=\linewidth]{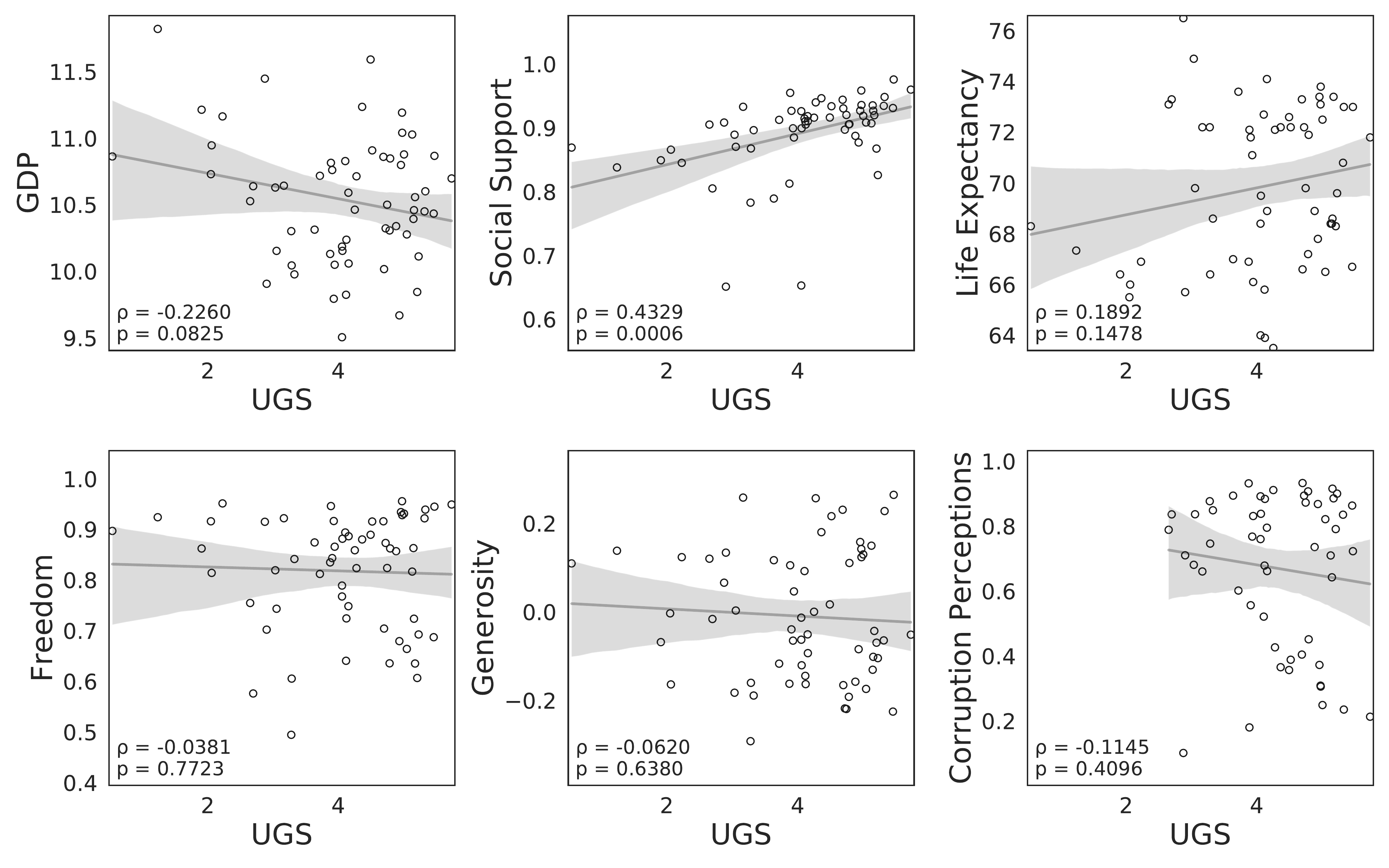}
\caption{Scatter plot between Green space(UGS) and variables in World Happiness Report. $\rho$ indicates the Pearson correlations.}
\label{fig:si_var}
\end{figure}

Here, we checked how our analyses fit into these six variables. The data of 6 variables are retrieved from the World Happiness Report, and we took a 3-year average. Figure \ref{fig:si_var} shows the scatter plots between UGS and six variables in the World Happiness Report. Note that the scatter plot between UGS and social support presents a relatively strong Pearson correlation of 0.4329, while other variables show no correlation with UGS. Therefore, we can suspect that the UGS is connected with the social support variable, which should be considered while constructing regression models.

\begin{table}[ht]
\centering
\small
\begin{tabular}{r|c c c|c c c}
\hline
& \multicolumn{3}{c}{Without corruption perceptions} & \multicolumn{3}{c}{With corruption perceptions}\\
\hline
	&	(1)	&	(2)	&	(3) &   (4) &   (5) &   (6)\\
\hline
GDP	&	0.5187\textsuperscript{***}	&	0.2388\textsuperscript{*}	&	0.2779\textsuperscript{*}    &   0.4694\textsuperscript{**}    &   0.1482  &   0.1512\\
 & (0.6508) & (0.5344) & (0.5991) & (0.9038) & (0.8178) & (0.8243)\\
UGS	&	0.1690\textsuperscript{***}	&	-	&	0.0339 & 0.1729\textsuperscript{**} & - & 0.0442\\
 & (0.2263) & & (0.2290) & (0.3255) & & (0.3018)\\
Social Support	&	-	&	5.1863\textsuperscript{***}	&	4.8452\textsuperscript{***} & - & 5.2457\textsuperscript{***} & 4.9787\textsuperscript{***}\\
 & & (3.6667) & (4.3514) & & (4.2056) & (4.6136)\\
Life Expectancy	&	0.0606\textsuperscript{***}	&	0.0556\textsuperscript{***}	&	0.0535\textsuperscript{***} & 0.0558\textsuperscript{**} & 0.0557\textsuperscript{**} & 0.0580\textsuperscript{**}\\
 & (0.0872) & (0.0733) & (0.0751) & (0.1115) & (0.0929) & (0.0950)\\
Freedom	&	2.4609\textsuperscript{***}	&	1.7472\textsuperscript{***}	&	1.7652\textsuperscript{***} & 2.2086\textsuperscript{***} & 1.5238\textsuperscript{**} & 1.5036\textsuperscript{**}\\
 & (2.6342) & (2.3277) & (2.3463) & (2.9002) & (2.5320) & 2.5549)\\
Generosity	& 0.6584 & 1.0520\textsuperscript{**}		& 1.0346\textsuperscript{**}	& 0.60031 & 1.0493\textsuperscript{**} & 1.0563\textsuperscript{**}\\
 & (2.0066) & (1.7401) & (1.7555) & (2.2309) & (1.9306) & (1.4819)\\
Corruption Perceptions	&	-	&	-	&	- & -0.3589 & -0.3109 & -0.3039\\
 & & & & (1.7389) & (8.6572) & (8.7725)\\
Const	&	-6.0323\textsuperscript{***}	&	-6.0873\textsuperscript{***}	&	-6.2005\textsuperscript{***} & -4.7584\textsuperscript{*} & -4.8097\textsuperscript{**} & -4.9458\textsuperscript{**}\\
 & (7.1740) & (6.0731) & (6.1610) & (10.2984) & (8.5672) & (8.7725)\\
\hline
Adjusted $R^2$	&	0.6753	&	0.7638	&	0.7609 & 0.6823 & 0.7730 & 0.7698\\
Observations & 59 & 59 & 59 & 54 & 54 & 54\\
\hline
\end{tabular}
\caption{\label{tab:si_var}Regression analysis of happiness with (1-3) 5 variables and (4-6) 6 variables in the World Happiness Report. We separated the models with corruption perceptions since few countries are missing data: Oman is excluded from the model (1-3), and Bahrain, Kuwait, Oman, Qatar, Saudi Arabia, and the United Arab Emirates are excluded from the model (4-6). \textsuperscript{***}p<0.01; \textsuperscript{**}p<0.05; \textsuperscript{*}p<0.1.}
\end{table}

Since the data for corruption perceptions is missing for six countries, and it seems to fail to explain happiness well for developed countries' data set, we checked the regression with and without corruption perception. The regression model (1) shows that UGS can explain happiness in place of social support, even although the adjusted R-square value is smaller compared to model (2), which includes social support. Furthermore, model (3), which includes both UGS and social support, shows that UGS loses its explainability while social support. The same result can be found in the model (4-6).




\newpage

\section{Moderated mediation model for regression}

The moderation and mediation technique provides a more complicated regression model, describing more detailed mechanisms behind regression. 

The mediation model describes the indirect effect of mediation variables described by the two-staged regression model. We applied the moderation model for log-GDP since we checked that the regression analyses for social support differed depending on the GDP value, which can be described with the cross term. We can set up the regression model as follows:
\begin{center}
\begin{equation}
    H = \beta_0 + \beta_1 M + \beta_2 S + \beta_3 SM
\end{equation}
\begin{equation}
    S = \beta_4 + \beta_5 G
\end{equation}
\end{center}

Now, we can validate the model with regression. The mediation model can be validated by comparing the multilinear regression model with its explanation of power. We will check whether green space affects happiness via social support.

\begin{table}[ht]
\centering
\small
\begin{tabular}{r|c c c c}
\hline
	&	(1)	&	(2)	&	(3) & (4)\\
\hline
log-GDP	&	1.1321\textsuperscript{***}	&	0.7168\textsuperscript{***}	&	0.7936\textsuperscript{***} & -4.1830\textsuperscript{**}\\
	&	(0.6193)	&	(0.5687)	&	(0.6336) & (6.566)\\
UGS	&	0.2457\textsuperscript{***}   &	      -  	&	0.0782 & -\\
	&	(0.2697)	&	        	&	(0.2863) &\\
Social Support	&	- &	6.3899\textsuperscript{***}	&	5.5731\textsuperscript{***} & -50.1512\textsuperscript{**}\\
	&				&	(4.4264)	&	(5.3363) & (75.625)\\
log-GDP:Social Support	&	-&	-	&-	 & 5.5583\textsuperscript{***}\\
	&				&		&	 & (7.423)\\
Const & -6.5695\textsuperscript{***}	&	-6.8962\textsuperscript{***}	&	-7.2953\textsuperscript{***} & 42.8656\textsuperscript{**}\\
	&	(6.8599)	&	(5.9001)	&	(6.0703) & (66.687)\\
\hline
Adjusted $R^2$	&	0.4912	&	0.6057	&	0.6071 & 0.6550\\
Observations & 59 & 59 & 59 & 59\\
\hline
\end{tabular}
\caption{\label{tab:si_mediation}Regression analysis for the moderated mediation model. Coefficient of GDP-Social Support represent cross term of GDP and social support. \textsuperscript{***}p<0.01; \textsuperscript{**}p<0.05; \textsuperscript{*}p<0.1.}
\end{table}

In Table \ref{tab:si_mediation}, model (1-3) describes the effect of UGS and social support on happiness. UGS and social support can explain happiness and GDP in the model (1) and (2). However, UGS loses its explainability when we include both UGS and social support in the model (3), which implies that UGS only indirectly affects happiness compared to social support. Note that our mediation model was valid for GDP, so the moderated mediation model would be more appropriate.

The moderation of the model can be validated by calculating the regression model with a cross-term. We check for moderation models in our consideration: moderation for green-social, social-happiness, and green-happiness. We find that the moderation effect emerges on the social-happiness relation with higher adjusted R-square and significantly low p-value (Table \ref{tab:si_mediation} model (4)). Therefore, we can conclude that the green space affects happiness through social support, and GDP moderates social support on happiness.

\section{Derivation of happiness equation}

How much we need green space to increase our happiness? Since our analyses are based on regression models, we can provide a numerical estimation of the required green space to increment happiness. Consider our final regression model:
\begin{center}
\begin{equation}
    H = \beta_0 + \beta_1 M + \beta_2 S + \beta_3 SM
\end{equation}
\begin{equation}
    S = \beta_4 + \beta_5 G
\end{equation}
\end{center}

where $H$ is the happiness score, $M$ is GDP per capita, $S$ is social support, and $G$ is UGS. If we substitute social support into the equation, we obtain the following equation.
\begin{center}
\begin{equation}
    H = \beta_0' + (\beta_1' + \beta_2' \ln{M})\ln{G} + \beta_3' \ln{M}
\end{equation}
\end{center}

If we assume that the value of GDP per capita stays the same, we can solve a fraction of green space change to increase a certain amount of happiness. We set the happiness score change to 0.0546, which is an average value for upgrading one rank.
\begin{center}
\begin{equation}
    \frac{G_f}{G_i} = \exp{\left( \frac{\Delta H}{\beta_1' + \beta_2' \ln{M}} \right)}
\end{equation}
\end{center}

\begin{figure}[ht]
\centering
\includegraphics[width=0.7\linewidth]{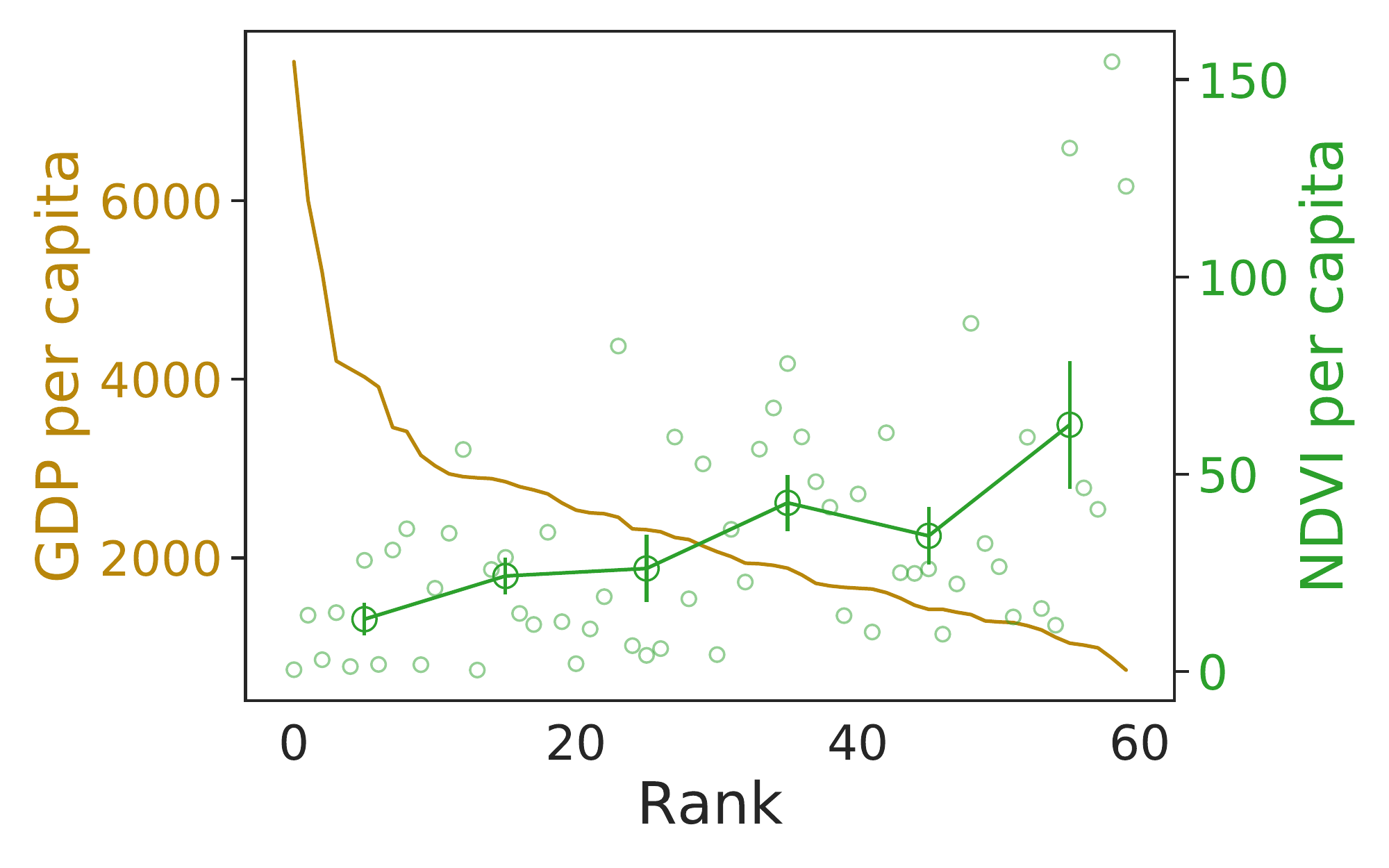}
\caption{Required GDP per capita (yellow) and NDVI per capita (green) to increase average amount of happiness for rank up.}
\label{fig:si_required}
\end{figure}

\begin{longtable}{r|ccc}
\hline
                 Country & Green Space [\%] & NDVI per capita & GDP per capita [dollar] \\
\hline
                Qatar &       14.50 &    0.4981 &  7556 \\
           Luxembourg &       16.00 &   14.3032 &  6004 \\
            Singapore &       17.10 &    3.0292 &  5199 \\
              Ireland &       19.04 &   14.9560 &  4205 \\
               Kuwait &       19.26 &    1.2950 &  4115 \\
               Norway &       19.49 &   28.2059 &  4026 \\
 United Arab Emirates &       19.79 &    1.8334 &  3914 \\
          Switzerland &       21.24 &   30.8228 &  3461 \\
        United States &       21.41 &   36.1908 &  3416 \\
         Saudi Arabia &       22.50 &    1.7651 &  3149 \\
          Netherlands &       23.05 &   21.1280 &  3032 \\
               Sweden &       23.52 &   35.0743 &  2941 \\
              Iceland &       23.68 &   56.2871 &  2909 \\
              Bahrain &       23.75 &    0.4071 &  2896 \\
            Australia &       23.80 &   25.8874 &  2888 \\
              Germany &       23.99 &   28.9519 &  2853 \\
              Austria &       24.31 &   14.7517 &  2797 \\
              Denmark &       24.53 &   11.9521 &  2761 \\
               Canada &       24.81 &   35.3155 &  2716 \\
              Belgium &       25.48 &   12.6587 &  2615 \\
                 Oman &       26.06 &    2.0204 &  2535 \\
               France &       26.29 &   10.8164 &  2504 \\
       United Kingdom &       26.37 &   18.9955 &  2495 \\
              Finland &       26.68 &   82.4975 &  2455 \\
                Malta &       27.81 &    6.5911 &  2325 \\
          South Korea &       27.90 &    4.1332 &  2315 \\
                Japan &       28.12 &    5.8425 &  2292 \\
          New Zealand &       28.75 &   59.4447 &  2229 \\
                Spain &       29.00 &   18.4593 &  2205 \\
                Italy &       29.80 &   52.6365 &  2133 \\
               Israel &       30.58 &    4.3248 &  2069 \\
       Czech Republic &       31.29 &   36.0511 &  2015 \\
  Trinidad and Tobago &       32.34 &   22.7089 &  1941 \\
             Slovakia &       32.44 &   56.3721 &  1934 \\
             Slovenia &       32.72 &   66.8141 &  1916 \\
            Lithuania &       33.22 &   78.0579 &  1885 \\
              Estonia &       34.51 &   59.4736 &  1810 \\
               Poland &       36.40 &   48.1184 &  1715 \\
             Portugal &       37.03 &   41.6695 &  1686 \\
             Malaysia &       37.40 &   14.1884 &  1670 \\
              Hungary &       37.63 &   45.0125 &  1660 \\
               Greece &       37.85 &   10.0445 &  1652 \\
               Latvia &       38.86 &   60.5196 &  1611 \\
               Russia &       40.60 &   25.0656 &  1549 \\
           Kazakhstan &       43.12 &   24.8977 &  1470 \\
               Panama &       44.84 &   26.0437 &  1424 \\
                Chile &       44.84 &    9.5004 &  1424 \\
              Romania &       46.17 &   22.2308 &  1391 \\
              Croatia &       47.31 &   88.2509 &  1365 \\
              Uruguay &       50.84 &   32.4731 &  1295 \\
            Mauritius &       51.53 &   26.5892 &  1283 \\
               Turkey &       51.91 &   13.8637 &  1276 \\
             Bulgaria &       54.05 &   59.3985 &  1241 \\
            Argentina &       57.43 &   15.9950 &  1193 \\
                 Iran &       64.68 &   11.7746 &  1112 \\
           Montenegro &       72.63 &  132.6388 &  1045 \\
              Belarus &       75.81 &   46.5536 &  1023 \\
           Costa Rica &       80.85 &   41.1266 &   993 \\
               Serbia &      111.22 &  154.5325 &   876 \\
              Albania &      213.00 &  122.9685 &   744 \\
\hline
\caption{\label{tab:si_required}Required green spaces and GDP for upgrading happiness.}
\end{longtable}